\documentclass[aps,pre,reprint,amsmath,amssymb,longbibliography,lengthcheck,showpacs,superscriptaddress,floatfix]{revtex4-1}
\usepackage{enumerate}
\usepackage{graphicx}
\usepackage{color}
\newcommand{\deri}[2]{{\displaystyle \frac{\partial #1 }{\partial #2 }}}

\begin{document}
\title{Structural, mechanical, and vibrational properties of particulate physical gels}
\author{Hideyuki Mizuno}
\email{hideyuki.mizuno@phys.c.u-tokyo.ac.jp} 
\affiliation{Graduate School of Arts and Sciences, The University of Tokyo, Tokyo 153-8902, Japan}
\author{Makoto Hachiya}
\affiliation{Graduate School of Arts and Sciences, The University of Tokyo, Tokyo 153-8902, Japan}
\author{Atsushi Ikeda}
\email{atsushi.ikeda@phys.c.u-tokyo.ac.jp} 
\affiliation{Graduate School of Arts and Sciences, The University of Tokyo, Tokyo 153-8902, Japan}
\affiliation{Research Center for Complex Systems Biology, Universal Biology Institute, The University of Tokyo, Tokyo 153-8902, Japan}
\date{\today}
%
\begin{abstract}
Our lives are surrounded by a rich assortment of disordered materials.
In particular, glasses are well known as dense, amorphous materials, whereas gels exist in low-density, disordered states.
Recent progress has provided a significant step forward in understanding the material properties of glasses, such as mechanical, vibrational, and transport properties.
In contrast, our understanding of particulate physical gels is still highly limited.
Here, using molecular dynamics simulations, we study a simple model of particulate physical gels, the Lennard-Jones (LJ) gels, and provide a comprehensive understanding of their structural, mechanical, and vibrational properties, all of which are markedly different from those of glasses.
First, the LJ gels show sparse, heterogeneous structures, and the length scale $\xi_s$ of the structures grows as the density is lowered.
Second, the gels are extremely soft, with both shear $G$ and bulk $K$ moduli being orders of magnitude smaller than those of glasses.
Third, many low-frequency vibrational modes are excited, which form a characteristic plateau with the onset frequency $\omega_\ast$ in the vibrational density of states.
Structural, mechanical, and vibrational properties, characterized by $\xi_s$, $G$, $K$, and $\omega_\ast$, respectively, show power-law scaling behaviors with the density, which establishes a close relationship between them.
Throughout the present work, we reveal that gels are multiscale, solid-state materials: (i) homogeneous elastic bodies at long lengths, (ii) heterogeneous elastic bodies with fractal structures at intermediate lengths, and (iii) amorphous structural bodies at short lengths.
\end{abstract}
\maketitle
%
\section{Introduction}~\label{sect:intro}
Disordered materials are ubiquitous in our daily lives.
Glasses are familiar examples at high density; examples include silicate glasses, metallic glasses, plastic materials, colloidal glasses, and so on~\cite{Phillips_1981,Berthier2011b,larson1999structure}.
In the glasses, particles are densely packed with disordered structures, and they behave as solids with rigidity.
On the other hand, gels are disordered materials at low density, which can be categorized into polymeric gels that are composed of polymers with cross-linking~\cite{larson1999structure,flory1953principles,de1979scaling} and particulate gels that are composed of particles such as colloids~\cite{larson1999structure,Zaccarelli2007,mewis2012colloidal,Lu2013,Ruiz_2021}.

In the gels, constituents are connected via bonds to form sparse heterogeneous network structures, such that they are of low density but still possess rigidity similar to solids.
According to the nature of the bonds, gels can be classified into chemical gels and physical gels~\cite{flory1953principles,Zaccarelli2007}.
In chemical gels, the bonds are chemical covalent bonds, which have virtually infinitely long lifetimes.
As a result, chemical gelation is an irreversible process that has been analyzed using the percolation theory for the bonded network~\cite{de1979scaling}.
Around the gelation point, the static and dynamic elastic moduli exhibit critical scaling laws, which have also been analyzed by means of percolation theory~\cite{de1979scaling}.

On the other hand, in physical gels, the bonds between particles originate from physical interactions, such as van der Waals interactions and depletion interactions~\cite{larson1999structure,mewis2012colloidal,Zaccarelli2007,Lu2013,Ruiz_2021}.
The strengths of these bonds are much weaker than those of covalent bonds~(typically of the order of thermal energy $k_B T$, where $T$ is temperature and $k_B$ is Boltzmann's constant), and formation of the physical gels is reversible and can be controlled by changing the physical conditions.
The present work concerns the material properties of particulate physical gels, with comparison to those of glasses.

Glasses are formed by lowering the temperature or increasing the density of particulate systems in their liquid states.
During these processes, the relaxation dynamics become increasingly sluggish, and the system loses the ability to relax into an equilibrium state at some point and falls into a nonequilibrium solid state, which is called the glass transition~\cite{Berthier2011b}.
The material properties of glasses are known to be widely different from those of crystalline solids~\cite{Phillips_1981}.
Their elastic moduli are smaller than those of their counterparts in crystalline states due to non-affine deformations~\cite{Tanguy_2002,Leonforte2005,Zaccone_2011,Mizuno_2013}.
Additionally, they show a characteristic excess of low-frequency vibrations, which is called the Boson peak~(BP)~\cite{Buchenau_1984,Yamamuro_1996,Mori_2020}.
The nature of vibrational states is changed at the BP frequency $\omega_\text{BP}$; below $\omega_\text{BP}$, phonon-like vibrational modes and quasi-localized vibrational~(QLV) modes are observed~\cite{Lerner_2016,Mizuno_2017,Shimada_2018,Wang_2019}, whereas above $\omega_\text{BP}$, vibrational modes show highly disordered vibrations~\cite{Silbert_2009,Mizuno_2017,Shimada_2018}, which are called anomalous modes~\cite{Wyart2_2005,Wyart_2006,Wyart_2005}.
Note that jammed states of short-ranged, soft repulsive particles play an important role in studies of these anomalous properties of glasses because this model exhibits the critical behavior of shear modulus due to the non-affine deformations~\cite{O'Hern2003,van_Hecke_2009} and abundance of the low-frequency vibrations, as evidenced in the flat vibrational density of state~(vDOS) $g(\omega) \propto \omega^0$ above the BP~\cite{Silbert_2005,Wyart_2005,Wyart2_2005,Wyart_2006}.

Particulate physical gels are formed by increasing the relative importance of attractive interactions between constitutive particles at low density, e.g., by adding salt to weaken the repulsive electrostatic interactions or by adding non-adsorbing polymers to strengthen the depletion interactions~\cite{larson1999structure,mewis2012colloidal,Zaccarelli2007,Lu2013,Ruiz_2021}.
When the density of particles is very low~($\varphi \lesssim 0.1$, where $\varphi$ is the packing fraction), attractive particles form fractal aggregates and ultimately become gels as the size of aggregates becomes macroscopic~\cite{fernandez2016fluids}.
This gelation process has been understood as diffusion-limited aggregation~(DLA), in which irreversible bonds are formed when two particles come into contact through diffusion processes~\cite{Witten_1983,Weitz_1985,Carpineti1992,Lu2013}. 
The elastic properties of DLA gels have been discussed in terms of percolation theory, as in the case of chemical gels, where the fractal dimension plays a key role in determining the scaling behaviors of the elastic moduli~\cite{Grant_1993,Krall_1998,Trappe2000,Prasad_2003,fernandez2016fluids}.

When the density is increased to a moderately low regime~($\varphi \gtrsim 0.1$), gelation takes place through \textit{arrested phase separation}~\cite{Zaccarelli2007,Lu2013}.
The equilibrium phase diagram of attractive particles includes gas and liquid phases, and the system undergoes gas-liquid phase separation when the attractive interaction is strong enough~\cite{Yamamoto_1994,Bailey_2007}.
However, when the attractive interaction is too strong, the equilibrium liquid phase cannot be realized because of the sluggish dynamics brought by the glass transition.
In this situation, the phase separation is interrupted so that the system forms the interconnected network structure of clusters of glasses, which is the gel state~\cite{Zaccarelli2007,Lu2013}.
At the level of the phase diagram, this scenario has been quantitatively verified by a combination of experimental and theoretical studies~\cite{Lu2008}.
However, the dynamics of this gelation process are not yet fully understood.
Numerical simulations~\cite{Testard_2011,Testard_2014} of the quenching of moderately low-density, attractive particles have demonstrated that such a system exhibits power-law growth of domain size in the short term followed by logarithmically slow long-term domain growth during the later stages.
For short-term behavior, the importance of dynamic asymmetry has been pointed out very recently~\cite{Tateno2021}.
For long-term behavior, similarity to the aging dynamics in glassy systems was pointed out~\cite{Testard_2014,oku2020phase}, but there is currently no quantitative understanding.
Furthermore, the mechanical properties of particulate physical gels are not yet understood and are under active debate~\cite{fernandez2016fluids}.
Various kinds of scenarios and explanations have been proposed, including jamming transition~\cite{Trappe_2008}, hierarchical arrest~\cite{Zaccone_2009,Zaccone2014}, locally favoured structures~\cite{Royall_2008}, local isostaticity~\cite{Hsiao16029,Tsurusawa2019}, rigidity percolation~\cite{Valadez2013}, and a correlated version of the rigidity percolation~\cite{Zhang2019}.

In the present work, we provide a comprehensive understanding of the structural, mechanical, and vibrational properties of particulate physical gels that are formed via arrested phase separation.
One particular focus is to discuss the similarity and difference between gels and glasses.
To achieve this, we focus on the simplest model: the zero-temperature quench of the system composed of Lennard-Jones~(LJ) particles.
Thanks to recent studies~\cite{Tanguy_2002,Leonforte2005,Mizuno_2013,Shimada_2018}, the mechanical and vibrational properties of LJ glasses are well understood, and we are now in a position to discuss the material properties of LJ gels based on comparing them to their glass counterparts.
Although the LJ particles might not be used for modeling colloidal systems, we believe that the understanding of the LJ gels provides a good starting point to understand the properties of colloidal gels, thus taking advantage of utilizing a well-established understanding of their glass counterparts.

\section{Methods}~\label{sect:methods}
\subsection{System description}~\label{subsect:system}
We have performed molecular dynamics~(MD) simulations on a model system that shows the glass transition at high density and gelation through arrested phase separation at lower density.
The system is composed of $N$ point particles in three-dimensional~($d=3$) space under periodic boundary conditions in all three directions.
Particles $i$ and $j$ interact through the LJ potential:
\begin{align}
\phi_\text{LJ}(r) = 4\epsilon \left[ \left( \frac{\sigma_{ij}}{r} \right)^{12} - \left( \frac{\sigma_{ij}}{r} \right)^{6} \right],
\end{align}
where $r$ is the distance between these two particles, $\sigma_{ij} = (\sigma_i + \sigma_j)/2$, and $\sigma_i$ is the size~(diameter) of particle $i$.
To avoid crystallization, we introduce a polydispersity in the distribution of particle sizes, as in Ref.~\cite{Leonforte2005}.
Specifically, the values of $\sigma_i$~($i=1,2,...,N$) are uniformly distributed in a range of $0.8 \sigma$ to $1.2 \sigma$.
The potential is cut off at $r = r_c = 3 \sigma$, where the potential and its first derivative are both made continuous as in~\cite{Shimada_2018}
\begin{align}
\phi(r) = \phi_\text{LJ}(r) - \phi_\text{LJ}(r_c) - (r-r_c) \frac{d\phi_\text{LJ}(r_c)}{dr}.
\end{align}
The mass $m$ is identical for all particles.
In the following, we employ $\epsilon$, $\sigma$, and $\tau = \sqrt{(m\sigma^2)/\epsilon}$ as units of energy, length, and time, respectively.
The temperature and the frequency are measured by units of $\epsilon/k_B$~($k_B$ is Boltzmann's constant) and $\tau^{-1} = \sqrt{\epsilon/(m\sigma^2)}$, respectively.

To study both the glass state and gel state, we vary the number density $\rho = N/V$~($V=L^3$ is the volume of the system, and $L$ is the linear dimension of the system) in a wide range of $\rho = 1.0,\ 0.7,\ 0.5,\ 0.3,\ 0.2$ and $0.1$, where the corresponding packing fractions are $\varphi = \left( \rho \pi/6 \right) \int_{0.8}^{1.2} \left( \sigma_i^3/0.4\right) d\sigma_i = 0.054,\ 0.11,\ 0.16,\ 0.27,\ 0.38$ and $0.54$, respectively.
We also vary the number of particles $N$ from $N = 10000$ to $640000$ to study finite system-size effects and access lower-frequency vibrational modes.

We first equilibrated the system in the normal liquid state at a temperature of $T=3.0$.
We then quenched the system to the zero-temperature state of $T=0$ by minimizing the system potential and bringing the system to the local potential minimum.
Here, we employ the steepest descent method~\cite{Press_2007} for minimization.
We numerically judge that the system settles down to a local potential minimum when the maximum value $f_\text{max}$ of the forces $\mathbf{F}_i$ that act on particles $i$~($i=1,2,...,N$) falls below $10^{-9}$.
Note that this protocol corresponds to an instantaneous quenching process with an infinite quenching rate.

In the following, we denote the $T=0$ configuration of particles~(or the inherent structure) as $\mathbf{r} =\left[ \mathbf{r}_{1},\mathbf{r}_{2},...,\mathbf{r}_{N} \right]$~($3N$-dimensional vector), where $\mathbf{r}_{i}$ is the position of particle~$i$.
As will be demonstrated in Sec.~\ref{sect:result:static}, we obtain the glass configuration at $\rho = 1.0$, while for the cases of $\rho = 0.7,\ 0.5,\ 0.3$ and $0.2$, we observe the gel configurations that are realized through arrested phase separation.

\subsection{Structural properties}~\label{subsect:method:structure}
We now obtained the $T=0$ configurations of particles, $\mathbf{r} =\left[ \mathbf{r}_{1},\mathbf{r}_{2},...,\mathbf{r}_{N} \right]$, at different densities $\rho$.
We first characterized the static structural properties of these configurations by calculating the radial distribution function $g(r)$ and the static structure factor $S(q)$~\cite{simpleliquid}.
We also measured the integrated value of $g(r)$, $N(r)$, as
\begin{align}
N(r)= \int_0^{r} 4\pi r'^2 \rho g(r') dr'.
\end{align}
Since $g(r)$ behaves as $g(r) \simeq 1$ at long distances $r \gg 1$, $N(r)$ converges to $\simeq  (4\pi \rho/3)r^3 \propto r^3$.
This behavior of $N(r) \propto r^3~(=r^d)$ indicates homogeneous media.
In contrast, $N(r) \propto r^{D_f}$ with the exponent of $D_f < 3~(=d)$ suggests sparse and heterogeneous structure or fractal-like structure, where $D_f$ is the fractal dimension~\cite{percolation}.

\subsection{Vibrational properties}~\label{subsect:method:vibration}
We next performed vibrational mode analysis on the $T=0$ configurations~\cite{Leibfried,Compute_vib}.
We solved the eigenvalue problem of the dynamical matrix $\mathbf{D}$~($3N \times 3N$ matrix) to obtain the eigenvalues $\lambda_k$ and the corresponding eigenvectors $\mathbf{e}^k=\left[ \mathbf{e}^{k}_1,\mathbf{e}^{k}_2,...,\mathbf{e}^{k}_N \right]$ for the modes $k=1,2,...,3N-3$ ($3$ zero-frequency, translational modes are removed).
The eigenvectors are orthonormalized as $\mathbf{e}^k \cdot \mathbf{e}^l = \sum_{i=1}^N \mathbf{e}_i^k \cdot \mathbf{e}_i^l = \delta_{k,l}$, where $\delta_{k,l}$ is the Kronecker delta function.

In the gels with $\rho =0.2$ and $0.3$, we found several vibrational modes with a zero eigenvalue, $\lambda_k=0$.
These modes with zero-energy cost emerge in clusters of particles that are isolated from the network structure of gels~(see Sec.~\ref{subsec:isolated} for details).
We denote the number of these zero-frequency modes as $N'$.
In analyses of the present work, these modes are disregarded.

We analyzed several different system sizes of $N=10000$ to $640000$
\footnote{
The largest system size $N_\text{max}$ used in the vibrational mode analysis depends on the density $\rho$; $N_\text{max} = 640000$ for $\rho=1.0$ and $0.7$, $320000$ for $\rho=0.5$, $160000$ for $\rho=0.3$, and $80000$ for $\rho=0.2$.
}.
We first calculated all the vibrational modes in the smallest system of $N=10000$.
We then calculated the low-frequency modes in the larger systems of $N=20000$ to $640000$.
Finally, the modes obtained from different system sizes were combined as a function of the frequency $\omega_k$.
We found that the results from different system sizes smoothly connect with each other, which enables us to extend the mode information to the lower-frequency regime~\cite{Mizuno_2017}.
In the figures of this paper, all data from different system sizes are presented together.

\subsubsection{Vibrational density of states}
From the dataset of eigenfrequencies, $\omega_k = \sqrt{\lambda_k}$ ($k=1,2,...,3N-3-N'$), the vibrational density of states~(vDOS) is calculated as 
\begin{equation} \label{equofvdos}
g(\omega) = \frac{1}{3N-3-N'} \sum_{k=1}^{3N-3-N'} \delta \left( \omega-\omega_k \right),
\end{equation}
where $\delta(x)$ is the Dirac delta function.
We have calculated the vDOSs by using different system sizes, which smoothly connect and give values in a wider~(lower) frequency regime~(see Fig.~S1 in the Supplementary Material~(SM)).

As will be described in Sec.~\ref{subsect:method:elastic}, we measured the bulk modulus $K$ and the shear modulus $G_0$.
From these values of elastic moduli, we calculated the Debye frequency $\omega_D$ and the Debye level $A_D$, as
\begin{equation}
\begin{aligned}
\omega_D &= \left[ \frac{ 18\pi^2 \rho }{ (c_L^{-3} + 2 c_T^{-3})} \right]^{1/3}, \\
A_D &= \frac{3}{\omega_D^3},
\end{aligned}
\end{equation}
where $c_L = \sqrt{(K+4G_0/3)/\rho}$ and $c_T = \sqrt{G_0/\rho}$ are longitudinal and transverse sound speeds, respectively.
We note that the ``minimum" value of shear moduli $G_0$ is employed for the calculation of these Debye values.

\subsubsection{Phonon order parameter}
The phonon order parameter $O_k$ evaluates the extent to which eigenvector $\mathbf{e}^k=\left[ \mathbf{e}^{k}_1,\mathbf{e}^{k}_2,...,\mathbf{e}^{k}_N \right]$ of the mode $k$ is similar to phonon vibrations~\cite{Mizuno_2017,Shimada_2018}.
We first define the displacement vectors of phonon vibrations as $\mathbf{e}^{\mathbf{q},\alpha}=\left[ \mathbf{e}^{\mathbf{q},\alpha}_1,\mathbf{e}^{\mathbf{q},\alpha}_2,...,\mathbf{e}^{\mathbf{q},\alpha}_N \right]$ with
\begin{equation}
\mathbf{e}^{\mathbf{q},\alpha}_i = \frac{1}{\sqrt{N}} \mathbf{s}_\alpha(\hat{\mathbf{q}}) \exp\left( {\text{i}\mathbf{q}\cdot \mathbf{r}_{i}} \right), \label{eq:phonon}
\end{equation}
where $\mathbf{q}$ represents the wave vector, $\hat{\mathbf{q}} = \mathbf{q}/\left| \mathbf{q} \right|$, and $\alpha$ denotes one longitudinal ($\alpha=L$) and two transverse ($\alpha=T_1,T_2$) phonon modes.
$\mathbf{s}_\alpha(\hat{\mathbf{q}})$ is a unit vector that represents the direction of polarization: $\mathbf{s}_{L}(\hat{\mathbf{q}}) = \hat{\mathbf{q}}$ (longitudinal) and $\mathbf{s}_{T_1}(\hat{\mathbf{q}}) \cdot \hat{\mathbf{q}}= \mathbf{s}_{T_2}(\hat{\mathbf{q}}) \cdot \hat{\mathbf{q}} = 0$ (transverse).

We then define the phonon order parameter $O_k$ as
\begin{equation} \label{phononorder}
\begin{aligned}
O_k & = \sum_{\mathbf{q},\alpha;\ O_k^{\mathbf{q},\alpha} \ge N_m/(3N-3-N')} O_k^{\mathbf{q},\alpha}, \\
O_k^{\mathbf{q},\alpha} & = \left| \mathbf{e}^{\mathbf{q},\alpha} \cdot \mathbf{e}^k \right|^2 = \left| \sum_{i=1}^N \mathbf{e}^{\mathbf{q},\alpha}_i \cdot \mathbf{e}^k_i \right|^2,
\end{aligned}
\end{equation}
where $N_m = 100$ is employed; however, we confirm that our results and conclusions do not depend on the choice of the value of $N_m$~\cite{Mizuno_2017,Shimada_2018}.
$O_k \approx 0$ indicates a mode considerably different from phonon vibrations, whereas finite values of $O^k >0$ indicate phonon-like vibrations.

Here, we note that $\mathbf{e}^{\mathbf{q},\alpha}$ does \textit{not} necessarily satisfy the orthonormal condition; $\mathbf{e}^{\mathbf{q},\alpha} \cdot \mathbf{e}^{\mathbf{q}',\alpha'} = \sum_{i=1}^N \mathbf{e}^{\mathbf{q},\alpha}_i \cdot \mathbf{e}^{\mathbf{q}',\alpha'}_i \neq \delta_{\mathbf{q},\mathbf{q}'} \delta_{\alpha,\alpha'}$, due to heterogeneous and amorphous structures, particularly for the gel configurations~(for the glass, the orthonormal condition is approximately satisfied~\cite{Mizuno_2017,Shimada_2018}).
Thus, it is possible that the value of $O_k$ exceeds one, $O_k >1$, which is indeed observed in the gels~(see Fig.~\ref{fig_character} and also Fig.~S4 in SM).

\subsubsection{Participation ratio}
The participation ratio $P_k$ quantitatively measures the extent of localization for each mode $k$, which has often been employed in many early works~\cite{Schober_1991,mazzacurati_1996,Taraskin_1999}.
Given the eigenvector $\mathbf{e}^k=\left[ \mathbf{e}^{k}_1,\mathbf{e}^{k}_2,...,\mathbf{e}^{k}_N \right]$, its participation ratio ${P}_k$ is calculated as
\begin{equation}~\label{eq:participation}
{P}_k \equiv \frac{1}{N} \left[ \sum_{i=1}^{N} \left( \mathbf{e}^k_i \cdot \mathbf{e}^k_i \right)^2 \right]^{-1}.
\end{equation}
${P}_k$ quantifies the fraction of particles that participate in the vibrations~($N P_k$ quantifies the number of participating particles)~\cite{Schober_1991,mazzacurati_1996,Taraskin_1999}.
As extreme cases, ${P}_k = 1$~($N P_k = N$) for an ideal mode in which all the constituent particles vibrate equally, and ${P}_k = 1/N \ll 1$~($N P_k = 1$) for an ideal mode involving only one particle.

\subsubsection{Vibrational energy}
We also calculated the vibrational energies of $\delta E_{k}^{\parallel}$ and $\delta E_{k}^{\perp}$ for each mode $k$~\cite{Mizuno_2017,Mizuno3_2016,Shimada2_2018}.
The vector of $\mathbf{e}_{ij}^k=\mathbf{e}_{i}^k-\mathbf{e}_{j}^k$ represents the vibrational motion between particles $i$ and $j$, which can be decomposed into the normal $\mathbf{e}_{ij}^{k \parallel}$ and tangential $\mathbf{e}_{ij}^{k \perp}$ vibrations with respect to the bond vector $\mathbf{n}_{ij}=(\mathbf{r}_{i}-\mathbf{r}_{j})/\left| \mathbf{r}_{i}-\mathbf{r}_{j} \right|$; $\mathbf{e}_{ij}^{k \parallel} = \left( \mathbf{e}^k_{ij} \cdot {\mathbf{n}_{ij}} \right) \mathbf{n}_{ij}$ and $\mathbf{e}_{ij}^{k \perp} = \mathbf{e}^k_{ij} - \left( \mathbf{e}^k_{ij} \cdot {\mathbf{n}_{ij}} \right) \mathbf{n}_{ij}$.
Accordingly, the vibrational energy $\delta E_k = \lambda_k/2 = \omega_k^2/2$ can be decomposed as
\begin{equation} \label{equofve}
\begin{aligned}
\delta E_k &= \sum_{\left< ij \right>} \left[ \frac{\phi''(r_{ij})}{2} \left( \mathbf{e}_{ij}^{k \parallel} \right)^2 + \frac{\phi'(r_{ij})}{2 r_{ij}} \left( \mathbf{e}_{ij}^{k \perp} \right)^2 \right], \\
&= \delta E_{k}^{\parallel} - \delta E_{k}^{\perp},
\end{aligned}
\end{equation}
where $\sum_{\left< ij \right>}$ denotes summation over all the interacting pairs of particles $\left< ij \right>$.
If the mode $k$ is phonon-like, then $\delta E_{k}^{\parallel}$ and $\delta E_{k}^{\perp}$ are both proportional to $\delta E_k \propto \omega_k^2$~\cite{Mizuno_2017}.
For the QLV modes and anomalous modes~(disordered vibrations) in glasses, the tangential energy $\delta E_{k}^{\perp}$ exhibits $\omega_k$-independent behavior, and $\delta E_{k}^{\perp} \propto \omega^0_k$~\cite{Mizuno3_2016,Mizuno_2017}.

\subsubsection{Spatial correlation of displacement field}
To study spatial correlations of the displacement field in the vibrational mode~$k$, we have calculated the correlation function $C_k(r)$~\cite{Silbert_2009}:
\begin{equation}
C_k(r=r_{ij}) = \left< \mathbf{e}^k_i (\mathbf{r}_i) \cdot \mathbf{e}^k_j(\mathbf{r}_j) \right>_{ij},
\end{equation}
where $\mathbf{e}^k_i(\mathbf{r}_i)$ of particle $i$ is denoted as a function of the position $\mathbf{r}_i$, and $r= r_{ij}= |\mathbf{r}_i-\mathbf{r}_j|$, and $\left< \right>_{ij}$ denotes the average over all the pairs of particles $ij$.
The phonon vibrations show long-range spatial correlations in $C_k(r)$.
Particularly, for the transverse phonons, negative correlations are observed due to the vortex-like displacement field~\cite{Mizuno2_2016}.
In contrast, anomalous modes in glasses show only short-range correlations with the order of particle size, representing disordered vibrations in nature~\cite{Silbert_2009}.

\subsection{Elastic moduli}~\label{subsect:method:elastic}
We finally analyzed the mechanical properties of the $T=0$ configurations, $\mathbf{r} =\left[ \mathbf{r}_{1},\mathbf{r}_{2},...,\mathbf{r}_{N} \right]$, at different densities $\rho$.
We calculated elastic moduli by using the fluctuation formulation developed based on linear response theory~\cite{lutsko_1989,Lemaitre_2006}.
Below, we write down the equations only and refer to Refs.~\cite{Compute_vib,Mizuno3_2016} for details of the explicit formulations.

The elastic modulus tensor $C_{\alpha \beta \gamma \delta}$~($\alpha,\beta,\gamma,\delta =x,y,z$) is composed of affine modulus $C^A_{\alpha \beta \gamma \delta}$ and non-affine modulus $C^N_{\alpha \beta \gamma \delta}$, as
\begin{equation}
C_{\alpha \beta \gamma \delta} = C^A_{\alpha \beta \gamma \delta} - C^N_{\alpha \beta \gamma \delta}.
\end{equation}
The affine $C^A_{\alpha \beta \gamma \delta}$ is formulated as
\begin{equation}
\begin{aligned}
C^A_{\alpha \beta \gamma \delta} &= C^B_{\alpha \beta \gamma \delta} + C^C_{\alpha \beta \gamma \delta}, \\
C^B_{\alpha \beta \gamma \delta} &= \frac{1}{V} \sum_{\left< ij \right>} \left( r_{ij}^2 \frac{d^2\phi(r_{ij})}{d{r_{ij}}^2} - {r_{ij}}\frac{d\phi(r_{ij})}{dr_{ij}} \right) \\
& \qquad \qquad \quad \times {n_{ij \alpha} n_{ij \beta} n_{ij \gamma} n_{ij \delta}}, \\
C^C_{\alpha \beta \gamma \delta} &= -\frac{1}{2} ( 2 \sigma_{\alpha \beta} \delta_{\gamma,\delta} -  \sigma_{\alpha \gamma} \delta_{\beta,\delta}-\sigma_{\alpha \delta} \delta_{\beta,\gamma} \\
& \qquad \qquad  -  \sigma_{\beta \gamma}\delta_{\alpha,\delta} - \sigma_{\beta \delta}\delta_{\alpha,\gamma} ),
\end{aligned}
\end{equation}
where $n_{ij \alpha}$ represents the bond vector, $\mathbf{n}_{ij}=(\mathbf{r}_{i}-\mathbf{r}_{j})/\left| \mathbf{r}_{i}-\mathbf{r}_{j} \right|$, and $\sigma_{\alpha \beta}$ is the stress tensor that is formulated as
\begin{equation}
\sigma_{\alpha \beta} = \frac{1}{V} \sum_{\left< ij \right>} \left( r_{ij} \frac{d\phi(r_{ij})}{dr_{ij}} \right) {n_{ij \alpha}n_{ij \beta}}.
\end{equation}
Note that $C^B_{\alpha \beta \gamma \delta}$ is the so-called Born term.
In addition, the non-affine $C^N_{\alpha \beta \gamma \delta}$ is formulated as
\begin{equation}
C_{\alpha \beta \gamma \delta}^{N} = \sum_{k=1}^{3N-3-N'} \frac{V}{\omega_k^2} \left( \sum_{i=1}^N \deri{\sigma_{\alpha \beta}}{\mathbf{r}_i} \cdot \mathbf{e}^{k}_i  \right)\left( \sum_{j=1}^N \deri{\sigma_{\gamma \delta}}{\mathbf{r}_j} \cdot \mathbf{e}^{k}_j \right),
\end{equation}
where we remind the reader that $N'$ zero-frequency modes appearing in gels are disregarded in the analysis, since they should make no contributions to the non-affine moduli.

From the modulus tensor $C_{\alpha \beta \gamma \delta}$, we have calculated two kinds of elastic moduli: bulk modulus $K$ for volume-changing bulk deformation and shear modulus $G$ for volume-preserving shear deformation, which are calculated as
\begin{widetext}
\begin{equation}
\begin{aligned}
K &= \frac{(C_{xxxx}+C_{yyyy}+C_{zzzz}+C_{xxyy}+C_{yyxx}+C_{xxzz}+C_{zzxx}+C_{yyzz}+C_{zzyy})}{9},\\
G_{1} & = \frac{(C_{xxxx}+C_{yyyy}-C_{xxyy}-C_{yyxx})}{4}, \\
G_{2} & = \frac{(C_{xxxx}+C_{yyyy}+4C_{zzzz}+C_{xxyy}+C_{yyxx}-2C_{xxzz}-2C_{zzxx}-2C_{yyzz}-2C_{zzyy})}{12}, \\
G_{3} &= C_{xyxy}, \qquad G_{4} = C_{xzxz}, \qquad G_{5} = C_{yzyz}.
\end{aligned}
\end{equation}
\end{widetext}
We note that for the shear modulus $G$, there are five independent values of $G_1,\ G_2,\ G_3,\ G_4,\ G_5$.
$G_1$ and $G_2$ are under \textit{pure} shear deformations (plane strain and
triaxial), and $G_3$, $G_4$, and $G_5$ are under \textit{simple} shear deformations.
These five values are the same in isotropic systems, whereas they can be different in anisotropic systems in general.
Indeed, we will observe anisotropic elastic properties with different values of shear modulus components in the gels~(see Fig.~S3 in SM).
In this work, we define two values of shear moduli: average value $G_\text{ave}$ and minimum value $G_0$, as
\begin{equation}
\begin{aligned}
G_\text{ave} &= \frac{(G_1 + G_2 + G_3 + G_4 + G_5)}{5}, \\
G_0 &= \text{min}(G_1,G_2,G_3,G_4,G_5).
\end{aligned}
\end{equation}

\begin{figure*}[t]
\centering
\includegraphics[width=0.95\textwidth]{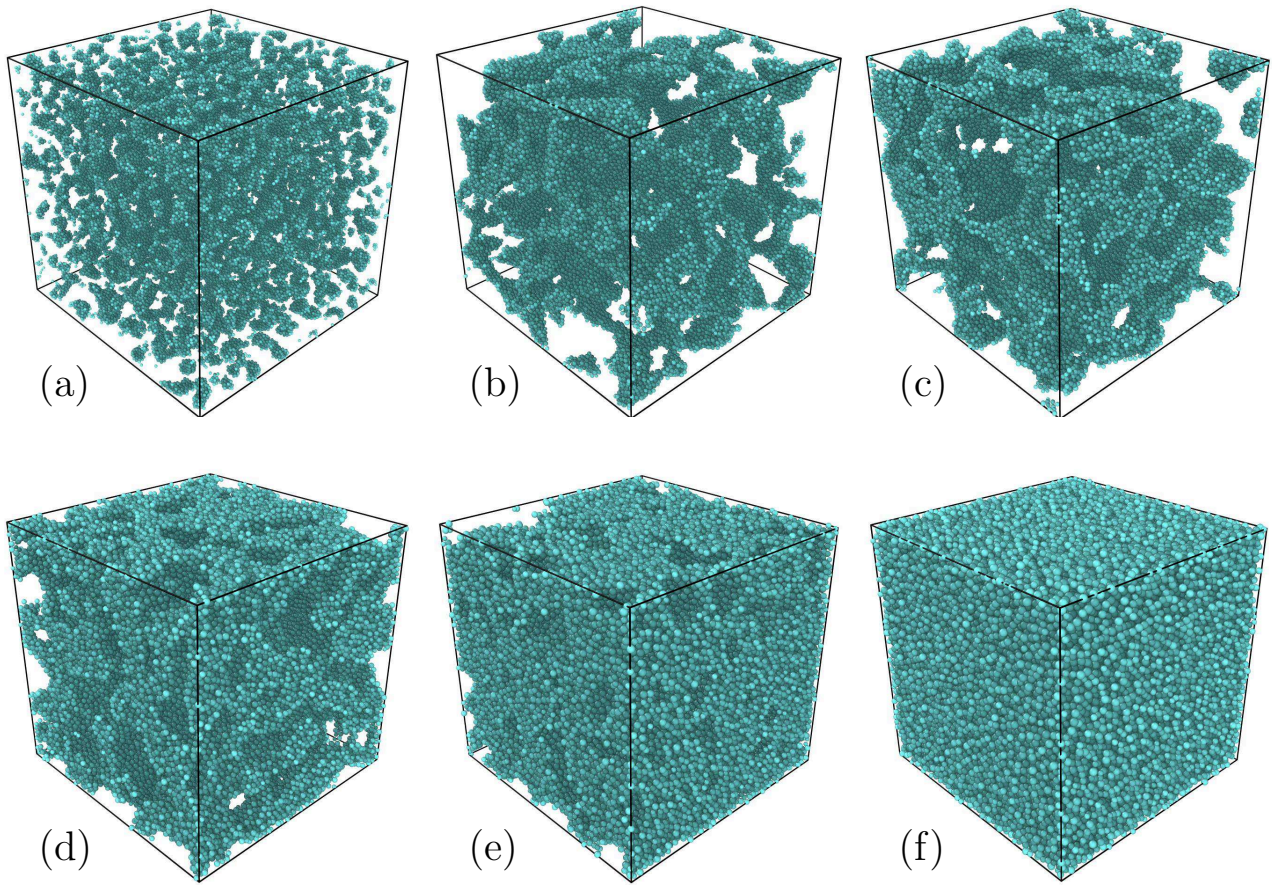}
\caption{\label{fig_visual}
{Snapshot of the system.}
The $T=0$ configurations are visualized for (a) $\rho = 0.1$~($\varphi = 0.054$), (b) $0.2$~($0.11$), (c) $0.3$~($0.16$), (d) $0.5$~($0.27$), (e) $0.7$~($0.38$), and (f) $1.0$~($0.54$).
The number of particles is $N=80000$, and the system length is (a) $L=92.8$, (b) $73.7$, (c) $64.4$, (d) $54.3$, (e) $48.5$, and (f) $43.1$.
}
\end{figure*}

\begin{figure}[t]
\centering
\includegraphics[width=0.475\textwidth]{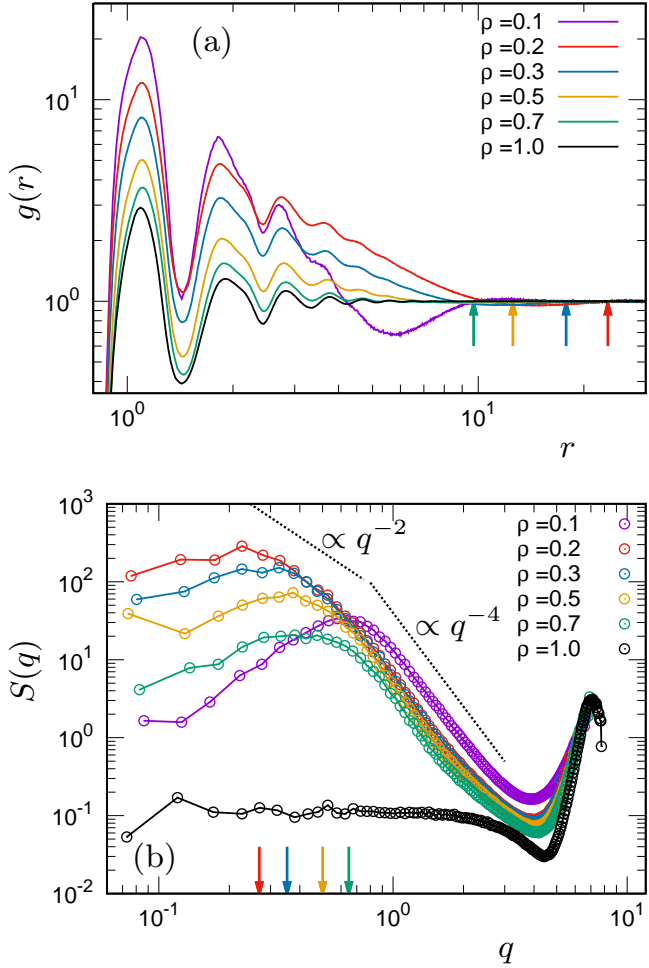}
\caption{\label{fig_rdf}
{Characterization of structural properties.}
(a) Radial distribution function $g(r)$ and (b) static structure factor $S(q)$ are plotted as functions of distance $r$ and wavenumber $q$, respectively, for the indicated values of $\rho$.
The arrows indicate the length $\xi_s$ in (a) and the wavenumber $q_s = 2\pi/\xi_s$ in (b) for $\rho = 0.2,\ 0.3,\ 0.5,\ 0.7$.
The dotted lines in (b) indicate the scaling of $S(q) \propto q^{-2}$ and the Porod law of $S(q) \propto q^{-4}$.
Note that $\xi_s$ is determined from the integrated value of $g(r)$, $N(r)$, as in Fig.~\ref{fig_intrdf}.
}
\end{figure}

\begin{figure}[t]
\centering
\includegraphics[width=0.475\textwidth]{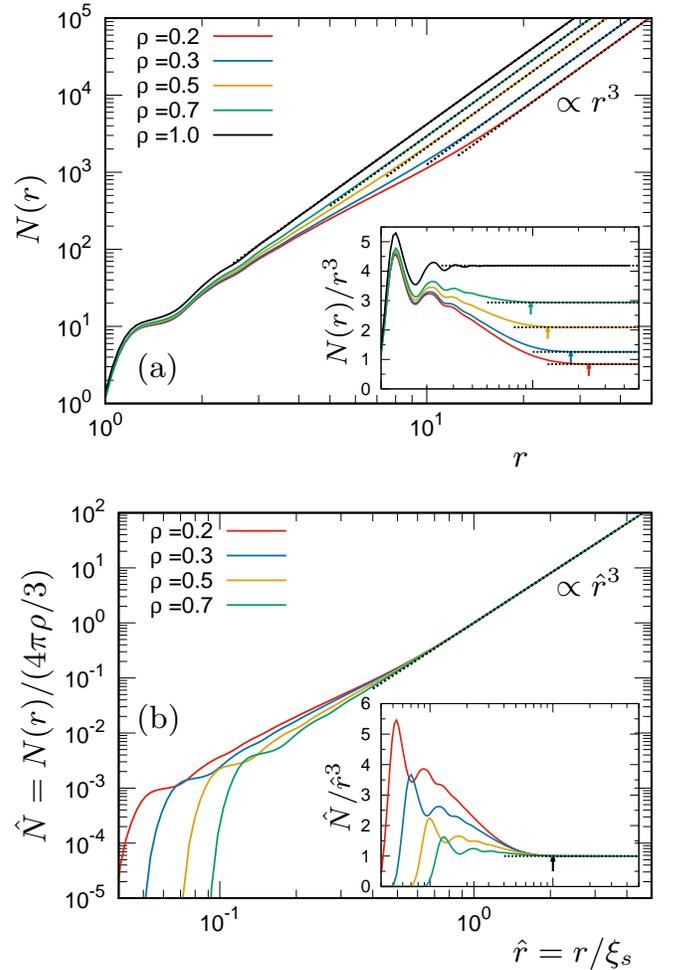}
\caption{\label{fig_intrdf}
{Integrated value of radial distribution function.}
(a) $N(r)$ and (b) normalized $\hat{N} = N(r)/(4\pi \rho/3)$ are plotted as functions of $r$ and $\hat{r} = r/\xi_s$, respectively, for the indicated values of $\rho$.
We also present data of $N(r)/r^3$ and $\hat{N}/\hat{r}^3$ in the insets of (a) and (b), respectively.
The length $\xi_s$ is determined as the distance above which $N(r)$ follows the $N(r) = (4\pi \rho/3) r^3 \propto r^3$ scaling law, which is indicated by dotted lines.
In the insets, arrows indicate values of length $\xi_s$.
}
\end{figure}

\begin{figure}[t]
\centering
\includegraphics[width=0.475\textwidth]{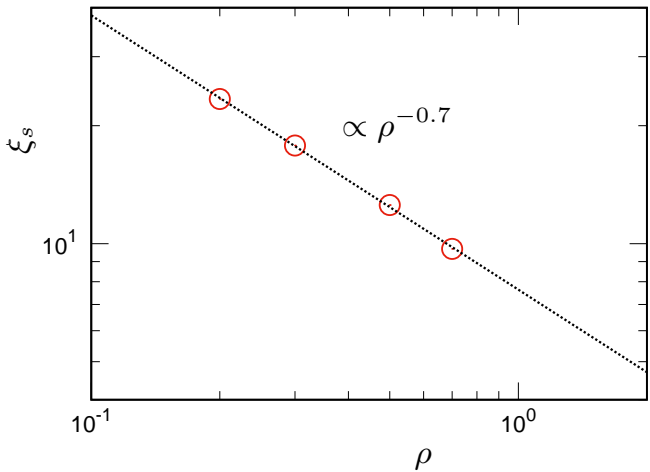}
\caption{\label{fig_ss_length}
{Characteristic length in static structure.}
The length $\xi_s$ is plotted as a function of $\rho$.
The line indicates a power-law scaling of $\xi_s \propto \rho^{-0.7}$.
}
\end{figure}

\begin{figure}[t]
\centering
\includegraphics[width=0.475\textwidth]{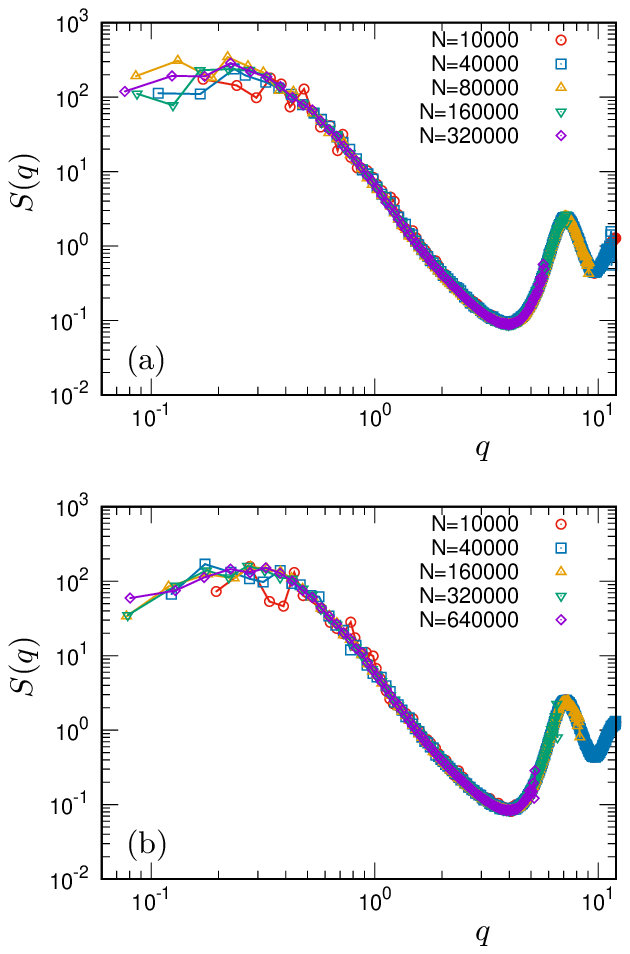}
\caption{\label{fig_ssf_size}
{Finite system-size effects on static structure factor.}
$S(q)$ is plotted as a function of $q$ for different system sizes of $N=10000$ to $640000$.
The density values are (a) $\rho=0.2$ and (b) $0.3$.
No noticeable differences are observed between different system sizes.
}
\end{figure}

\section{Structural properties}~\label{sect:result:static}
\subsection{Snapshot of system}
Figure~\ref{fig_visual} visualizes the $T=0$ configurations in three-dimensional space for densities of $\rho = 0.1,\ 0.2,\ 0.3,\ 0.5,\ 0.7$, and $1.0$.
For the highest $\rho = 1.0$, we obtain the glass configuration.
In the glass, particles are tightly packed, and the glass structure is disordered but spatially homogeneous.
As the density is lowered to $\rho = 0.7$ to $0.2$, the system undergoes phase separation between gas and glass during the rapid quenching process~\cite{Testard_2011,Testard_2014}, which results in sparse and heterogeneous configurations at $T=0$.
We note that $\rho = 0.2$ to $0.7$ at $T=0$ lies within the spinodal line of phase separation between gas and liquid~\cite{Asano_2012}.
Particularly, for the case of $\rho = 0.2$, we observe a rather heterogeneous, network-like structure where clusters of particles are connected with each other.
Finally, for the lowest $\rho = 0.1$, the density is so low that the network structure is sufficiently broken down to be disconnected.
In this state, isolated clusters of particles are distributed in space.

\subsection{Characterization of structural properties}~\label{subsect:result:grsq}
To characterize structural properties quantitatively, we present the radial distribution function $g(r)$ in Fig.~\ref{fig_rdf}(a) and the static structure factor $S(q)$ in Fig.~\ref{fig_rdf}(b).
At $\rho = 1.0$, we observe characteristics of the glasses: $g(r)$ shows spatial correlation only at particle sizes of $r \sim 1$, and accordingly, $S(q)$ shows density fluctuations only at approximately $q \sim 2\pi$.
As the density decreases towards $\rho = 0.2$, $g(r)$ shows the long-range correlations, and accordingly significant enhancement of $S(q)$ emerges at small wavenumbers $q$.
These results of $g(r)$ and $S(q)$ quantify the heterogeneous structures visualized in Fig.~\ref{fig_visual}.

As the density is lowered from $\rho =1.0$ to $0.7$, a precipitous shift is observed in $S(q)$ in Fig.~\ref{fig_rdf}(b).
This behavior is attributed to spinodal decomposition~(between gas and glass phases) at $T=0$.
Previous works demonstrated that the glass system at $T=0$ experiences spinodal instability when lowering the density~\cite{Sastry_2000,Altabet_2016,Shimada_2020}.
This observation is consistent with that the $T=0$ systems of $\rho = 0.2$ to $0.7$ are the gel states that are generated through arrested phase separation.

When examining $S(q)$ of $\rho = 0.2$ to $0.7$~(gels) in more detail, we find that at $q \sim 0.8$ to $3$, $S(q)$ follows the so-called Porod law~\cite{Bray_1994}, $S(q) \propto q^4$, indicating that the density field has sharp interfaces.
This confirms that tightly packed particles form clusters and that such dense clusters are connected with each other to form a heterogeneous, network-like structure.
Since $S(q)$ values at large $q \sim 2\pi$~(order of particle size) are similar to those of $\rho=1.0$~(glass), clusters are in glass-solid states.
At $q \lesssim 0.8$ and $\rho =0.2$ and $0.3$, we observe that $S(q)$ roughly follows $\propto q^{-2}$, and this scaling regime extends to the smaller $q$ with $\rho$ declining from $0.3$ to $0.2$.
This result implies the existence of a fractal character of dimension $D_f \simeq 2$~\cite{percolation,Nakayama_1994}.
For the lowest $\rho = 0.1$, the connections of these clusters disappear, and they become isolated in space, as visualized in Fig.~\ref{fig_visual}.
The values of $S(q)$ at small $q$ are reduced, which means that the distribution of these isolated clusters tends to be homogeneous.

\subsection{Characteristic length}~\label{subsect:result:length}
We next present the integrated value of $g(r)$, $N(r)$, in Fig.~\ref{fig_intrdf}~(see also Fig.~S2 in SM).
At long distances, it is observed that $N(r) \simeq  (4\pi \rho/3)r^3 \propto r^3~(=r^d)$, which crosses over to $N(r) \propto r^{D_f}$ with an exponent of $D_f < 3$ at intermediate distances for the gel configurations of $\rho = 0.2$ to $0.7$.
The exponent $D_f$ decreases from $\simeq 3$ to $\simeq 2$ with decreasing density $\rho$: $D_f=2.9,\ 2.7,\ 2.4,\ 2.2$ for $\rho = 0.7,\ 0.5,\ 0.3,\ 0.2$~(see Fig.~S2 in SM).
These results indicate that a fractal-like structure is developed at intermediate distances in the gels~\cite{percolation,Nakayama_1994}.
The values of $D_f \simeq 2$ for $\rho =0.2$ and $0.3$ are consistent with the scaling of the static structure factor, $S(q) \propto q^{-2}$, observed in Fig.~\ref{fig_rdf}(b).

For the gels with $\rho = 0.2$ to $0.7$, we define a characteristic length $\xi_s$ as the onset distance of $N(r) \simeq  (4\pi \rho/3)r^3 \propto r^3$.
At $r<\xi_s$, $N(r) \propto r^{D_f}$ with $D_f <3$, indicating sparseness and heterogeneities, whereas at $r > \xi_s$, $N(r) \propto r^3$, indicating that heterogeneities are coarse-grained and that the structure becomes homogeneous.
For reference, Figure~\ref{fig_rdf} plots the arrow values of $\xi_s$ in (a) and the corresponding wavenumber $q_s = 2\pi/\xi_s$ in (b).
We observe that $g(r)$ converges to $1$ without oscillatory behavior at $r \ge \xi_s$, and correspondingly, $S(q)$ converges to flat behavior at $q \le q_s$.

Finally, we plot the length $\xi_s$ as a function of $\rho$ in Fig.~\ref{fig_ss_length}.
The length grows with decreasing $\rho$, representing growing heterogeneities that are consistent with snapshots in Fig.~\ref{fig_visual}.
Remarkably, the length follows a power-law scaling of
\begin{equation}
\xi_s \propto \rho^{-0.7}. \label{eq.length}
\end{equation}
This observation implies the existence of a critical phenomenon with the critical point at $\rho = 0$.
We will discuss this point at the end of this paper in Sec.~\ref{sect:conclusion}, after presenting all of the results regarding mechanical and vibrational properties.

\subsection{Finite size effects}~\label{subsect:result:fs}
Finally, we mention the system-size effects on static structures.
Figure~\ref{fig_ssf_size} plots the static structure factor $S(q)$ for different system sizes of $N=10000$ to $640000$.
No noticeable differences are recognized between different system sizes.
We therefore conclude that the heterogeneous structures of the gels are not affected by finite-system-size effects.
This result establishes that the characteristic length $\xi_s$ remains finite even at the thermodynamic limit of $N \to \infty$ in the $T=0$ quenching of systems; namely, the phase separation in this protocol is indeed arrested, and the LJ systems form well-defined gel states.

\begin{figure}[t]
\centering
\includegraphics[width=0.475\textwidth]{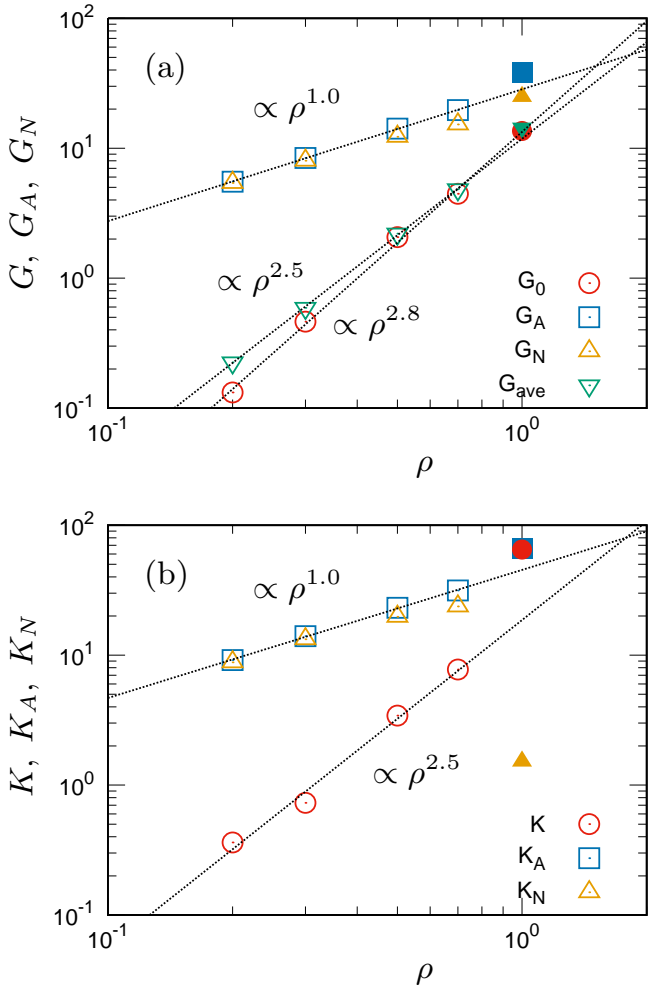}
\caption{\label{fig_moduli}
{Elastic moduli.}
(a) $G$ and (b) $K$ are plotted as functions of $\rho$.
For $G$, two values are presented: the average value $G_\text{ave} = (G_1 + G_2 + G_3 + G_4 + G_5)/5$, and the minimum value $G_0 = \text{min}(G_1,G_2,G_3,G_4,G_5)$.
We also plot the affine $G_A,\ K_A$ and non-affine $G_N,\ K_N$ components.
Note that $G_A$ and $G_N$ correspond to the components of $G_0$.
Closed symbols represent data of the glass with $\rho=1.0$.
The lines indicate power-law scalings of $G_\text{ave} \propto \rho^{2.5}$, $G_0 \propto \rho^{2.8}$, and $G_A,\ G_N \propto \rho^{1.0}$ in (a), and $K \propto \rho^{2.5}$ and $K_A,\ K_N \propto \rho^{1.0}$ in (b).
}
\end{figure}

\section{Mechanical properties}~\label{sect:result:elastic}
Figure~\ref{fig_moduli} plots the shear modulus $G$~($G_\text{ave}$, $G_0$) in (a) and the bulk modulus $K$ in (b) as functions of $\rho$.
We first remark that the glass with $\rho = 1.0$ shows isotropic shear elasticity with nearly the same values of $G_1$ to $G_5$, and so $G_\text{ave} \simeq G_0$.
However, in the gels, the shear moduli become anisotropic with different values of $G_1$ to $G_5$, as shown in Fig.~S3 of SM.
In Fig.~\ref{fig_moduli}, we find differences between $G_\text{ave}$ and $G_0$ with lowering $\rho$.
We expect that this anisotropy originates from the sparse, heterogeneous structures of gels.

As $\rho$ is lowered, $G$ and $K$ decrease significantly and similarly.
Comparing $\rho=0.2$~(gel) and $1.0$~(glass), the moduli of gels are orders of magnitude smaller than those of glasses; $G_\text{ave}=0.22,\ G_0 = 0.13,\ K=0.36$ at $\rho=0.2$ and $G_\text{ave} \approx G_0 = 14,\ K=65$ at $\rho=1.0$.
This result demonstrates that gels are extremely soft under both shear and bulk deformations.
In Fig.~\ref{fig_moduli}, the affine and non-affine components are also presented.
For amorphous systems, the non-affine moduli are important components to determine the elastic moduli~\cite{Tanguy_2002,Leonforte2005,Zaccone_2011,Mizuno_2013}.
As $\rho$ is lowered, the affine $G_A$ and $K_A$ are nearly canceled by the non-affine $G_N$ and $K_N$, respectively, which results in very small values of $G=G_A-G_N$ and $K=K_A-K_N$.

Note that the non-affine component $K_N$ in bulk modulus is contrasting between gels and glasses.
In the glasses, $K_N$ is negligible compared to the affine $K_A$, $K_N \ll K_A$, and $K \approx K_A$~\cite{Leonforte2005,Mizuno_2013}.
Since particles are homogeneously packed in the glasses, their displacements under isotropic bulk deformation are approximately along the affine deformation.
In contrast, $K_N$ is comparable to $K_A$ in the gels.
Due to the sparse and heterogeneous structures, non-affine motions of particles are induced in the gels even under isotropic deformation.

It is remarkable that $G$ and $K$ of the gels are observed to follow power-law scalings of
\begin{equation}
G_\text{ave} \propto \rho^{2.5}, \quad G_0 \propto \rho^{2.8}, \quad K \propto \rho^{2.5}. \label{eq.moduli}
\end{equation}
The affine components also follow power-law scalings of $G_A \propto \rho^{1.0},\ K_A \propto \rho^{1.0}$, which is simply because the system becomes sparse with lowering density.
To eliminate effects from reductions in affine components, we can consider the scalings of
\begin{equation}
\frac{G_\text{ave}}{G_A} \propto \rho^{1.5}, \quad \frac{G_0}{G_A} \propto \rho^{1.8}, \quad \frac{K}{K_A} \propto \rho^{1.5},
\end{equation}
which are rather nontrivial scalings due to the non-affine deformations.
This observation again implies the existence of the critical point at $\rho = 0$, where both $G/G_A$ and $K/K_A$ vanish and the system loses solidity, which will be discussed at the end of this paper in Sec.~\ref{sect:conclusion}.

\begin{figure}[t]
\centering
\includegraphics[width=0.475\textwidth]{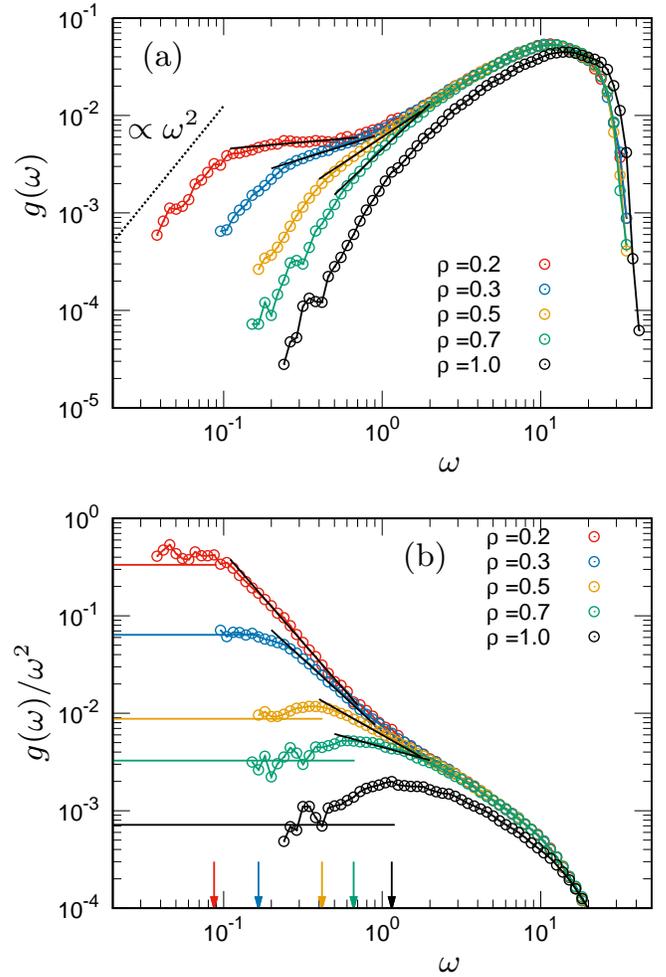}
\caption{\label{fig_vdos}
{Vibrational density of states.}
(a) $g(\omega)$ and (b) $g(\omega)/\omega^2$ are plotted as functions of $\omega$ for the densities of $\rho = 0.2,\ 0.3,\ 0.5,\ 0.7$~(gels) and $\rho = 1.0$~(glass).
In panel (b), the horizontal lines indicate the Debye level $A_D$.
Additionally, the arrows in (b) indicate the characteristic frequency $\omega_\ast$ at which $g(\omega)/\omega^2$ takes a maximum value for $\rho = 0.5,\ 0.7,\ 1.0$ or converges to $A_D$ for $\rho = 0.2,\ 0.3$.
Black solid lines to the data of gels with $\rho = 0.2$ to $0.7$ indicate $g(\omega) \propto \omega^{\tilde{d}-1}$ with the spectral dimension $\tilde{d}$.
Values of $\tilde{d}$ are presented in Table~\ref{table1}.
}
\end{figure}

\begin{figure}[t]
\centering
\includegraphics[width=0.475\textwidth]{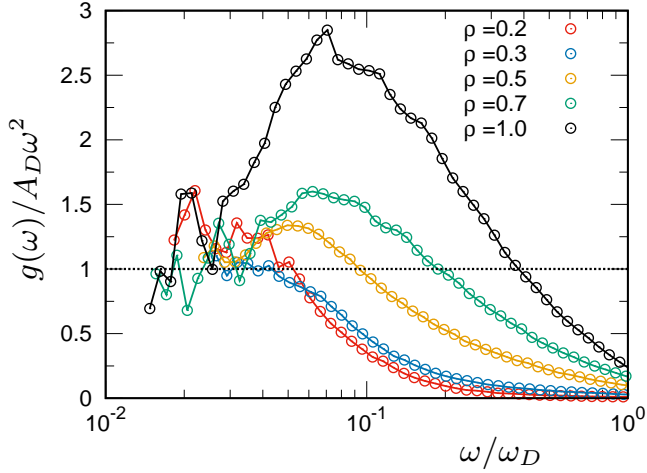}
\caption{\label{fig_scalevdos}
{Scaled vibrational density of states.}
$g(\omega)/A_D\omega^2$ is plotted as a function of $\omega/\omega_D$ for the densities of $\rho = 0.2,\ 0.3,\ 0.5,\ 0.7$~(gels) and $1.0$~(glass).
The dotted line corresponds to the Debye vDOS $g(\omega) = A_D \omega^2$.
The peak height is reduced towards $1$ by lowering $\rho$, and no apparent peak exists at $\rho = 0.2$ and $0.3$.
}
\end{figure}

\begin{figure*}[t]
\centering
\includegraphics[width=0.95\textwidth]{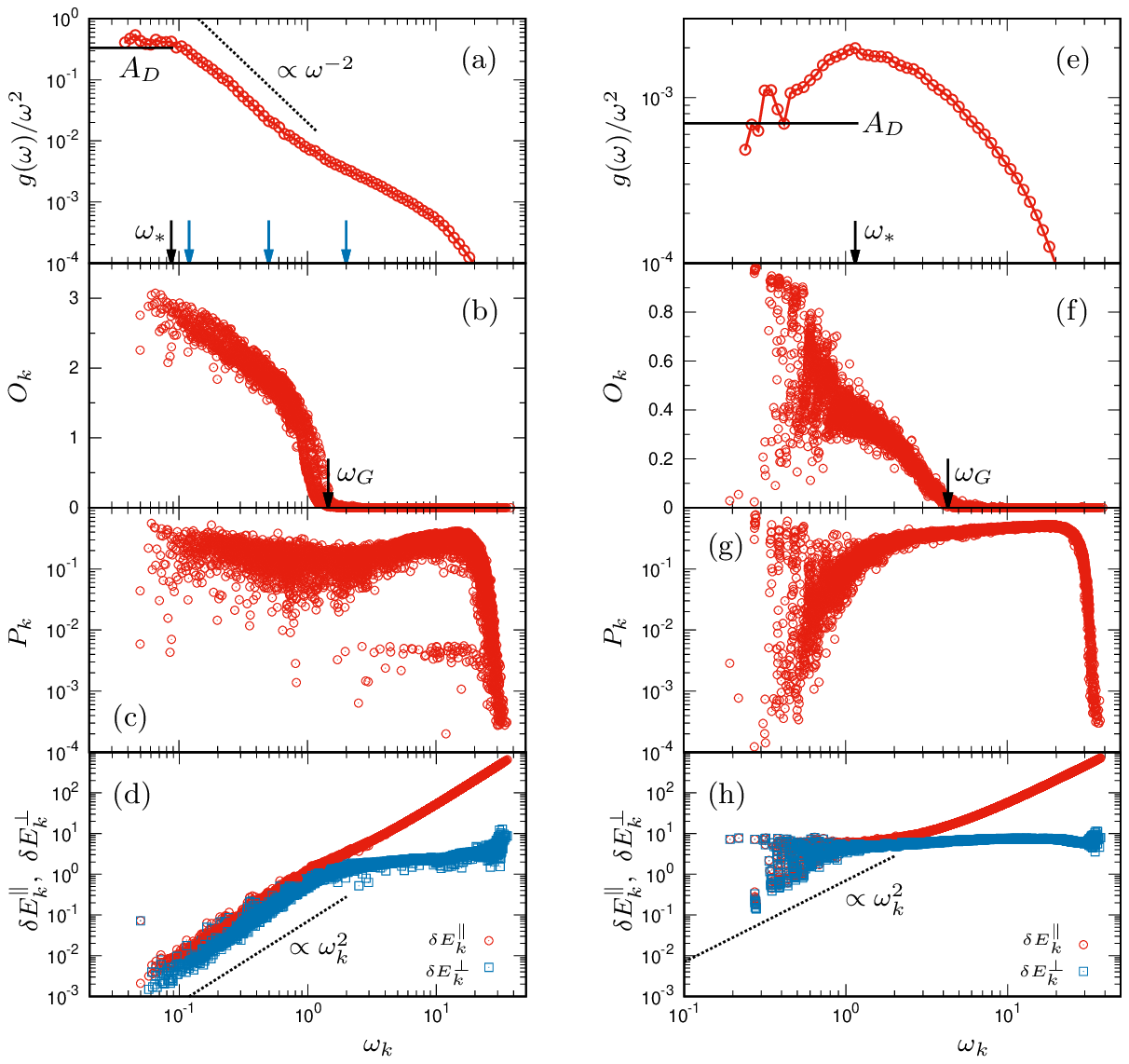}
\caption{\label{fig_character}
{Characterization of vibrational modes.}
Plots of (a,e) $g(\omega)/\omega^2$ and (b,f) $O_k$, (c,g) $P_k$, (d,h) $\delta E_k^{\parallel}$, and $\delta E_k^{\perp}$ of each mode $k$ as functions of $\omega_k$.
The left panels of (a) to (d) are data of the gel with $\rho=0.2$, and the right panels of (e) to (h) are those of the glass with $\rho=1.0$.
The data of $g(\omega)/\omega^2$ are the same as those presented in Fig.~\ref{fig_vdos}(b).
The horizontal lines in panels~(a,e) present the Debye level $A_D$.
The black arrows in (a,e) indicate the frequency $\omega_\ast$, whereas those in~(b,f) indicate $\omega_G$ at which $O_k$ becomes zero, $O_k \approx 0$.
The blue arrows in panel~(a) indicate values of frequencies that correspond to three visualized modes in Fig.~\ref{fig_visvsgel}.
}
\end{figure*}

\begin{figure*}[t]
\centering
\includegraphics[width=0.95\textwidth]{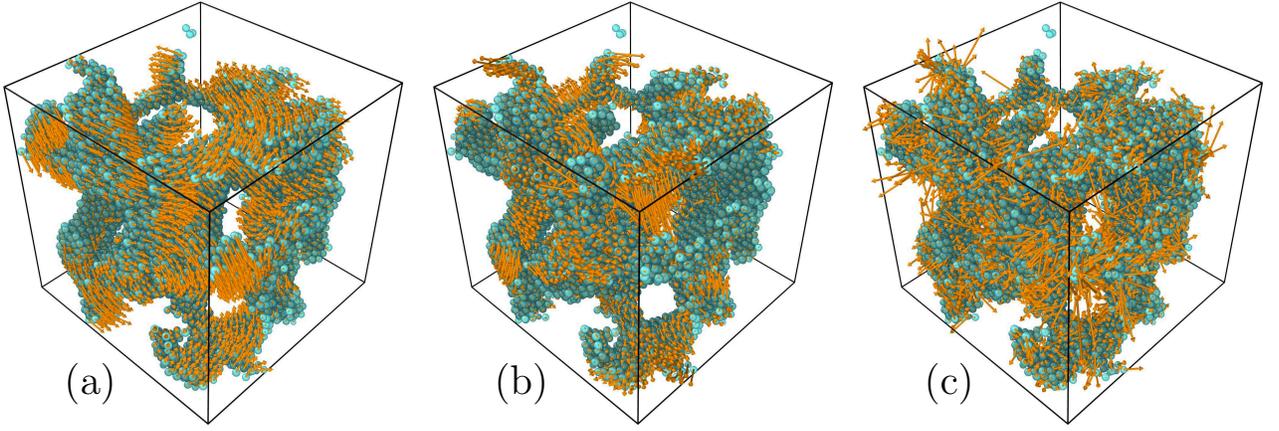}
\caption{\label{fig_visvsgel}
{Visualization of vibrational modes in the gel with $\rho =0.2$.}
Vibrational states of (a) $\omega_k = 0.12$, $O_k=2.8$, $P_k=0.66$, (b) $\omega_k = 0.50$, $O_k=1.9$, $P_k=0.28$, and (c) $\omega_k = 2.0$, $O_k=0.014$, $P_k=0.21$ are visualized.
The number of particles is $N=10000$, and the system length is $L=36.8$.
In the figures, $200 \times \mathbf{e}^k_i$~($i=1,2,...,N$) are plotted by arrows.
Values of frequencies corresponding to three visualized modes are indicated in Fig.~\ref{fig_character}(a).
}
\end{figure*}

\begin{figure}[t]
\centering
\includegraphics[width=0.475\textwidth]{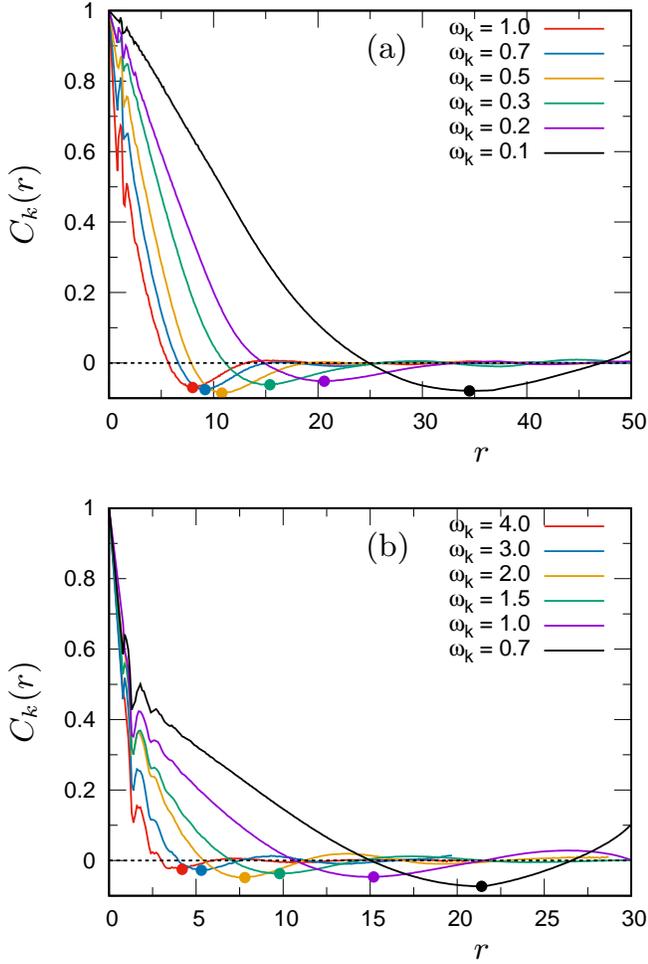}
\caption{\label{fig_corf}
{Spatial correlation function for vibrational modes.}
$C_k(r)$ is plotted as a function of $r$ for (a) $\rho =0.2$~(gel) and (b) $\rho=1.0$~(glass) and for the modes with indicated values of $\omega_k$ below $\omega_G$.
$C_k(r)$ shows nonmonotonic dependence on $r$, crossing the zero value of $C_k(r) =0$ and exhibiting a negative correlation.
The length $\xi_k$ is extracted as the distance $r=\xi_k$ at which $C_k(r)$ takes a (negative) minimum value, which is indicated by closed circles.
}
\end{figure}

\begin{figure}[t]
\centering
\includegraphics[width=0.475\textwidth]{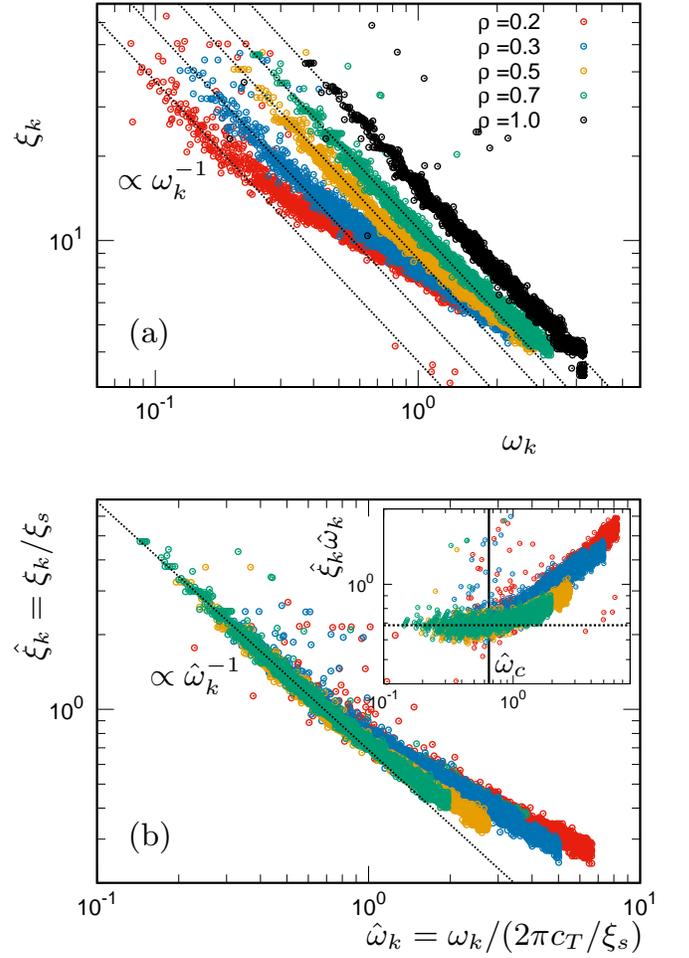}
\caption{\label{fig_length}
{Correlation length of vibrational modes.}
(a) $\xi_k$ is plotted as a function of $\omega_k$ for densities of $\rho = 0.2$, $0.3$, $0.5$, $0.7$, and $1.0$.
(b) Scaled $\hat{\xi}_k = \xi_k/\xi_s$ is plotted by scaled $\hat{\omega}_k = \omega_k/(c_T q_s) = \omega_c /(2\pi c_T / \xi_s)$~(where $q_s = 2\pi/\xi_s$) for the gels with $\rho = 0.2$, $0.3$, $0.5$, and $0.7$.
The inset of (b) plots $\hat{\xi}_k \hat{\omega}_k$ versus $\hat{\omega}_k$.
The lines indicate $\propto \omega_k^{-1}$ in (a) and $\propto \hat{\omega}_k^{-1}$ in (b).
$\hat{\xi}_k \propto \hat{\omega}_k^{-1}$ collapses for different densities below the crossover frequency of $\hat{\omega}_c = \omega_c/(c_T q_s) = \omega_c / (2\pi c_T / \xi_s)$.
From the inset of (b), we obtain $\omega_c = 0.65 c_T q_s= 1.3 \pi c_T / \xi_s$.
}
\end{figure}

\begin{figure}[t]
\centering
\includegraphics[width=0.475\textwidth]{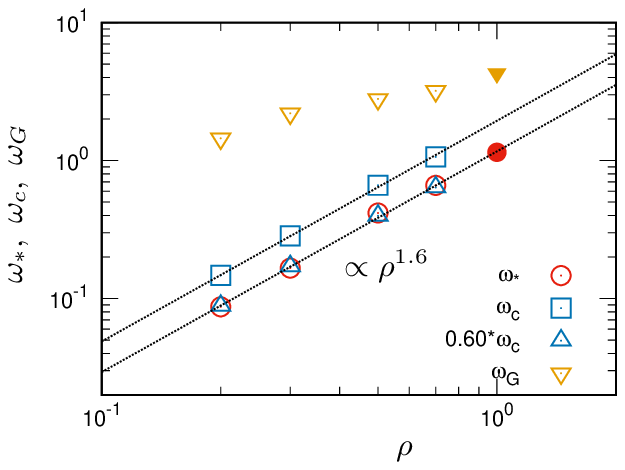}
\caption{\label{fig_frequency}
{Characteristic frequencies.}
$\omega_\ast$, $\omega_c~(= 0.65 c_T q_s= 1.3 \pi c_T / \xi_s)$, and $\omega_G$ are plotted as functions of $\rho$.
Closed symbols represent data of the glass with $\rho=1.0$.
$\omega_\ast \approx 0.60 \omega_c$ is observed for the gels with $\rho = 0.2$ to $0.7$.
The lines demonstrate $\omega_\ast \approx 0.60 \omega_c \approx 0.78 \pi c_T / \xi_s \propto \rho^{1.6}$.
}
\end{figure}

\section{Vibrational properties}~\label{sect:result:vibration}
We next present vibrational properties of the gels.
We first calculated the vDOS $g(\omega)$ and then characterized each vibrational mode in terms of phonon order parameter $O_k$, participation ratio $P_k$, and vibrational energies $\delta E_{k}^{\parallel}$ and $\delta E_{k}^{\perp}$.
We also measured the spatial correlations $C_k(r)$ for each vibrational mode.

\subsection{Vibrational density of states}
Figure~\ref{fig_vdos}(a) shows the vDOS $g(\omega)$ for the densities of $\rho = 0.2,\ 0.3,\ 0.5,\ 0.7$~(gels) and $1.0$~(glass).
It is remarkable that as the density is lowered in the gels, a characteristic plateau develops in the low-frequency regime.
This observation demonstrates that many low-frequency modes emerge in the gels, which form the plateau in the vDOS.
With further detailed analyses, the plateau regime is characterized in terms of $g(\omega) \propto \omega^{\tilde{d}-1}$ with the spectral dimension $\tilde{d}$~\cite{Nakayama_1994}.
This point will be described later in Sec.~\ref{subsec:spectral}.

In addition, we plot the reduced vDOS $g(\omega)/\omega^2$ in Fig.~\ref{fig_vdos}(b).
In the glass with $\rho=1.0$, we observe a clear peak above the Debye level $A_D$, which is BP~\cite{Buchenau_1984,Yamamuro_1996,Mori_2020}.
BP is also observed in the gels with $\rho = 0.5$ and $0.7$.
However, as $\rho$ decreases, the BP is reduced, and at $\rho =0.2$ and $0.3$, the BP disappears, so that $g(\omega)/\omega^2$ smoothly converges to $A_D$.
We also plot $g(\omega)/(A_D \omega^2)$ in Fig.~\ref{fig_scalevdos}, where the peak height in $g(\omega)/(A_D \omega^2)$ is significantly reduced towards $1$ with lowering $\rho$, and then no apparent peak exists at $\rho = 0.2$ and $0.3$.
We can therefore conclude that the BP properties are markedly different between glasses and gels: BP is developed in glasses, whereas it is reduced and even absent in gels.
In this work, we define the frequency $\omega_\ast$ at which $g(\omega)/\omega^2$ takes a maximum value for $\rho = 0.5,\ 0.7,\ 1.0$ or converges to $A_D$ for $\rho = 0.2,\ 0.3$.
The $\omega_\ast$ thus corresponds to the BP frequency or onset frequency of the plateau in $g(\omega)$.

\subsection{Characterization of vibrational states}
We next characterize each vibrational mode $k$ in terms of the phonon order parameter $O^k$, participation ratio $P^k$, and vibrational energies $\delta E^{k \parallel}$ and $\delta E^{k \perp}$, which are presented in Fig.~\ref{fig_character}.
In this figure, we show data of the gel with $\rho =0.2$ and those of the glass with $\rho=1.0$ for comparison.
Additional data for gels with $\rho=0.3$ and $0.5$ are available in Fig.~S4 of SM.

Let us first examine the glass with $\rho = 1.0$ in the right panels of Fig.~\ref{fig_character}.
In addition to $\omega_\ast$, we define another characteristic frequency $\omega_G$ at which $O_k$ becomes zero, $O_k \approx 0$: i.e., the character of phonon vibrations completely disappears at this frequency.
In the glass, $\omega_\ast$ and $\omega_G$ are of the same order of magnitude, $\omega_\ast \sim \omega_G$.
Above $\omega_\ast \sim \omega_G$, we observe the anomalous modes~\cite{Wyart2_2005,Wyart_2006,Wyart_2005} which are disordered with $O_k \approx 0$ and extended with large $P_k$.
The tangential energy follows $\delta E_{k}^{\perp} \propto \omega_k^0$, and the normal energy $\delta E_{k}^{\parallel} \propto \omega_k^2$ becomes orders of magnitude larger than $\delta E_{k}^{\perp}$; $\delta E_{k}^{\perp} \ll \delta E_{k}^{\parallel} \approx \delta E_{k} \propto \omega_k^2$.
On the other hand, below $\omega_\ast \sim \omega_G$, phonon-like modes with large values of $O_k$ and $P_k$ and QLV modes with small $O_k$ and $P_k$ are observed.
The phonon modes show $\delta E^{k \parallel}, \delta E^{k \perp} \propto \omega_k^2$, whereas the QLV modes follow $\delta E^{k \parallel}, \delta E^{k \perp} \propto \omega_k^0$.
Thus, the character of the modes is qualitatively changed at $\omega_\ast$~($\sim \omega_G$); phonon vibrations exist due to elastic-body character below $\omega_\ast$, whereas above $\omega_\ast$, rather disordered vibrations emerge due to structurally amorphous character.
These observations of glasses were already reported in Refs.~\cite{Mizuno3_2016,Mizuno_2017,Shimada_2018}
\footnote{
In Ref.~\cite{Mizuno_2017}, we denote $\omega_\ast$ (the BP frequency) as $\omega_\text{BP}$, while $\omega_G$ is denoted as $\omega_\ast$.
}.

We now turn our attention to the gel with $\rho =0.2$ in the left panels of Fig.~\ref{fig_character}.
We also visualize the vibrational states in Fig.~\ref{fig_visvsgel} for three representative modes whose frequencies are indicated by blue arrows in panel (a) of Fig.~\ref{fig_character}.
First, when examining $P_k$~(and comparing it to that of the glass), we do not recognize the apparent existence of the QLV modes below $\omega_\ast$.
We find some modes with low values of $P_k$.
However, their values of $O_k$ are not small and $\delta E^{k \parallel}, \delta E^{k \perp} \propto \omega_k^2$.
These observations indicate that these modes with low $P_k$ show phonon-like vibrations that are different in nature from the QLV modes in glasses.
Although QLV modes might be detected in larger system sizes, we can conclude that QLV modes are substantially suppressed or even absent in the gels.

We next examine the frequency of $\omega_G$ in the gel.
Interestingly, the plateau of vDOS terminates at approximately $\omega_G$.
Above $\omega_G$, vibrations are very similar as anomalous modes in glasses, which show $O_k \approx 0$~(disordered), large $P_k$~(extended), and $\delta E_{k}^{\perp} \propto \omega_k^0 \ll \delta E_{k}^{\parallel} \approx \delta E_{k} \propto \omega_k^2$~(see also Fig.~\ref{fig_visvsgel}(c) for visualization).
Note that some modes with small $P_k$ are observed above $\omega_G$, which will be discussed later in Sec.~\ref{subsec:isolated}.
On the other hand, below $\omega_G$, vibrations become phonon-like with increasing $O_k$, and vibrational energies follow the scaling of $\delta E^{k \parallel}, \delta E^{k \perp} \propto \omega_k^2$~(see also Figs.~\ref{fig_visvsgel}(a) and (b) for visualization).
These phonon-like modes form the characteristic plateau in the vDOS of gel at $\omega < \omega_G$.

We emphasize again that $\omega_\ast$ and $\omega_G$ of glasses are equivalent scales with the same order of magnitude.
In contrast, in the gel with $\rho=0.2$, these two frequencies are on different scales with different orders of magnitude, $\omega_\ast \ll \omega_G$.
$\omega_\ast$ and $\omega_G$ provide the onset and end frequencies of the plateau in vDOS, respectively.
As will be shown in Sec.~\ref{subsec:frequency} and Fig.~\ref{fig_frequency}, these two frequencies depend on $\rho$ differently, such that their differences become wider with lowering $\rho$.
The character of vibrational modes is changed at these frequencies.
The higher-frequency $\omega_G$ is the boundary between the elastic-body and amorphous-structure characteristics.
Above $\omega_G$, disordered vibrations emerge due to amorphous-structure character, whereas below $\omega_G$, phonon-like vibrations associated with elastic-body character persist.
The nature of the phonon vibrations below $\omega_G$ is changed at the lower-frequency $\omega_\ast~(\ll \omega_G)$.
The phonon vibrations above $\omega_\ast$ correspond to sparse, elastic bodies with heterogeneous network-like structures, whereas those below $\omega_\ast$ are associated with homogeneous elastic bodies.
This crossover behavior at $\omega_\ast$ will be described next in Sec.~\ref{subsec:correlation}.

\subsection{Spatial correlation of displacement field}~\label{subsec:correlation}
To further examine the nature of phonon-like vibrations below $\omega_G$ in the gels, we present the spatial correlation function $C_k(r)$ for modes $k$ of $\omega_k < \omega_G$ in Fig.~\ref{fig_corf}.
We can see that for both the gel and the glass, $C_k(r)$ first decreases, takes a negative minimum, and then converges towards zero value with oscillation.
This behavior represents typical vibrational states of transverse phonons~\cite{Mizuno2_2016}, which show vortex structure, as visually recognized in Figs.~\ref{fig_visvsgel}(a) and (b).
Note that the shear modulus $G_0$ is smaller than half the value of the bulk modulus $K$ in both the gels and glass, as plotted in Fig.~\ref{fig_moduli}.
This result indicates that transverse phonon vibrations are dominant over longitudinal phonon vibrations in the low-frequency regime, which is consistent with the observations in Figs.~\ref{fig_visvsgel} and~\ref{fig_corf}.
We then define the length $\xi_k$ for each vibrational mode~(of $\omega_k \le \omega_G$) as the distance $r = \xi_k$ at which $C_k(r)$ first takes a (negative) minimum value, as indicated by closed circles in Fig.~\ref{fig_corf}.
$\xi_k$ measures the length of the vortex structure, which corresponds to half of the wavelength of (transverse) phonon vibrations.

Figure~\ref{fig_length}(a) plots $\xi_k$ versus $\omega_k$ for gels with $\rho=0.2$ to $0.7$ and glass with $\rho=1.0$.
The glass shows $\xi_k \propto \omega_k^{-1}$ dependence for the entire range of $\omega_k \le \omega_G$.
The exponent ``$-1$" indicates the behavior of phonons propagating through homogeneous elastic media~\cite{Ashcroft_1976,Nakayama_1994}.
As seen in Sec.~\ref{sect:result:static}, the structure of the glass is rather homogeneous, which is consistent with the exponent ``$-1$" and $\xi_k \propto \omega_k^{-1}$.

On the other hand, the gels show crossover behavior at some frequency $\omega_k = \omega_c$: below $\omega_c$, $\xi_k \propto \omega_k^{-1}$ is observed, whereas $\xi_k \propto \omega_k^{-1/a}$~($\omega_k \propto \xi_k^{-a}$) with $1/a < 1$~($a>1$) is observed above $\omega_c$~(see also Fig.~S5 in SM).
The exponent of $1/a < 1$~($a>1$) picks up properties of phonon vibrations in heterogeneous elastic media~\cite{Nakayama_1994}.
The value of $1/a$ decreases~($a$ increases) with lowering of the density: $1/a=0.88,\ 0.77,\ 0.63,\ 0.52$~($a=1.1,\ 1.3,\ 1.6,\ 1.9$) for $\rho=0.7,\ 0.5,\ 0.3,\ 0.2$, respectively~(see Fig.~S5 in SM).
As $\rho$ is lowered, the structure becomes sparser and more heterogeneous, which causes the smaller value of $1/a$~(the larger value of $a$).
We therefore conclude that the properties of phonon vibrations are changed at $\omega_k = \omega_c$: below $\omega_c$, phonon vibrations occur through homogeneous elastic media, while above $\omega_c$, vibrations occur in heterogeneous elastic media.

We now scale data of $\xi_k$ versus $\omega_k$ of gels by using the length $\xi_s$~(of static structure) and the frequency $c_T q_s = 2\pi c_T/\xi_s$, where $c_T=\sqrt{G_0/\rho}$ is the speed of transverse phonons and $q_s = 2\pi /\xi_s$ is the characteristic wavenumber corresponding to $\xi_s$.
Here, we employ the minimum value of shear moduli, $G_0 = \text{min}(G_1,G_2,G_3,G_4,G_5)$, for the calculation of $c_T$, which is the most closely related to the low-frequency vibrations.
Figure~\ref{fig_length}(b) plots $\hat{\xi}_k = \xi_k/\xi_s$ versus $\hat{\omega}_k = \omega_k/(c_T q_s) = \omega_k/(2\pi c_T/ \xi_s)$.
Remarkably, the crossover point at $\hat{\omega}_c = \omega_c/(c_T q_s) = \omega_c/(2\pi c_T/ \xi_s)$ and data of $\hat{\omega}_k < \hat{\omega}_c$ collapse onto a single curve of $\hat{\xi}_k \propto \hat{\omega}_k^{-1}$ for different densities $\rho$.
From the inset of Fig.~\ref{fig_length}(b), we can determine $\omega_c$ as $\omega_c \approx 0.65 c_T q_s = 1.3 \pi c_T/\xi_s$.
These results indicate that the crossover behavior is determined by the length $\xi_s$ and the (transverse) phonon speed $c_T$ or the shear modulus $G_0$.
For wavelengths longer than $\xi_s$, phonons do \textit{not} experience sparsenesses and heterogeneities in the gels: they thus behave as if they propagate in the homogeneous media.
On the other hand, for wavelengths shorter than $\xi_s$, phonons encounter sparsenesses and heterogeneities.
We therefore conclude that phonon vibrations below $\omega_G$ are controlled by the static structural properties and the shear rigidity.

\subsection{Characteristic frequencies}~\label{subsec:frequency}
To date, we obtain three characteristic frequencies of $\omega_\ast$, $\omega_G$, and $\omega_c \approx 0.65 c_T q_s= 1.3 \pi c_T/\xi_s$, which are explicitly plotted as functions of $\rho$ in Fig.~\ref{fig_frequency}.
For the glass with $\rho=1.0$, $\omega_\ast$ and $\omega_G$ are on the same order of magnitude, as already mentioned.
As the density $\rho$ is lowered to the gel states, both frequencies decrease; however, $\omega_G$ is rather insensitive to $\rho$, such that the values of $\omega_G$ stay on the same order of magnitude as that of the glass~($\rho=1.0$).
In contrast, $\omega_\ast$ decreases significantly and becomes orders of magnitude smaller than $\omega_G$.
As a result, $\omega_\ast$ and $\omega_G$ reach different scales in the gels.

Turning attention to $\omega_\ast$ and $\omega_c$ of the gels, it is remarkable that both frequencies depend on $\rho$ in the same manner, as $\omega_\ast \approx 0.60\omega_c$.
This result means that $\omega_\ast$~(BP frequency or onset frequency of plateau) has the same physical meaning as $\omega_c$: below $\omega_\ast$, phonon vibrations propagate through homogeneous media, whereas above $\omega_\ast$, they are associated with sparse and heterogeneous media.

In addition, $\omega_\ast \approx 0.60\omega_c$ provides an important relationship between the frequency $\omega_\ast$, the length scale in the static structure $\xi_s$, and the transverse sound speed $c_T$ or the shear modulus $G_0$ as
\begin{align}
\omega_\ast \approx 0.60 \omega_c \approx 0.78 \pi \frac{c_T}{\xi_s} \approx 0.78\pi \frac{\sqrt{G_0}}{\xi_s \sqrt{\rho}}. \label{eq:bpf}
\end{align}
Substituting the scalings of $\xi_s \propto \rho^{-0.7}$~[in Eq.~(\ref{eq.length}) and Fig.~\ref{fig_ss_length}] and $G_0 \propto \rho^{2.8}$~[in Eq.~(\ref{eq.moduli}) and Fig.~\ref{fig_moduli}] into $\omega_\ast$ in Eq.~(\ref{eq:bpf}), we obtain
\begin{equation}
\omega_\ast \propto \omega_c \propto \rho^{1.6},
\end{equation}
which is indeed consistent with the observation of Fig.~\ref{fig_frequency}.
From Eq.~(\ref{eq:bpf}), we can conclude that the onset regime of the plateau is composed of phonon-like modes with wavelengths comparable to the length scale of the heterogeneous structure.

\begin{table}[t]
\caption{\label{table1}
Summary of values of fractal dimension $D_f$, exponent $a$ of dispersion curve, and spectral dimension $\tilde{d}$ in the gel states of $\rho = 0.2$ to $0.7$.} 
\centering
\renewcommand{\arraystretch}{1.5}
\begin{tabular}{c|c|c|c|c|c|c}
\hline
\hline
$\rho$ &\ $0.2$ \ &\ $0.3$ \ &\ $0.5$ \ &\ $0.7$ \ &\ \text{Def.}\ &\ \text{Ref.} \\
\hline
\hline
$D_f$ &\ 2.2 \ &\ 2.4 \ &\ 2.7 \ &\ 2.9 \ &\ $N(r) \propto r^{D_f}$ \ &\ Fig.~S2 of SM \\
\hline
$a$ &\ 1.9 \ &\ 1.6 \ &\ 1.3 \ &\ 1.1 \ &\ $\omega_k \propto \xi_k^{-a}$ \ &\ Fig.~S5 of SM \\
\hline
$\tilde{d}=D_f/a \ $\ &\ 1.1 \ &\ 1.5 \ &\ 2.1 \ &\ 2.6 \ &\ $g(\omega) \propto \omega^{\tilde{d}-1}$ \ & Fig.~\ref{fig_vdos} \\
\hline
\hline
\end{tabular}
\end{table}

\subsection{Spectral dimension of vDOS}~\label{subsec:spectral}
Here, we discuss the spectral dimension of vDOS for the gel states of $\rho = 0.2$ to $0.7$.
In Sec.~\ref{subsect:result:length} and Figs.~\ref{fig_intrdf} and~S2 of SM, we have obtained
\begin{equation}
N(r)
\left\{ \begin{aligned}
& \propto r^3~(=r^d) & (r > \xi_s), \\
& \propto r^{D_f} & (r < \xi_s),
\end{aligned} \right.~\label{eq:fractal}
\end{equation}
which indicates the existence of a fractal-like structure with fractal dimension $D_f < 3$ at $r<\xi_s$.
In addition, in Sec.~\ref{subsec:correlation} and Figs.~\ref{fig_length} and~S5 of SM, we have shown
\begin{equation}
\omega_k
\left\{ \begin{aligned}
&\propto \xi_k^{-1} & (\xi_k > \xi_s\ \&\ \omega_k < \omega_c),\\
&\propto \xi_k^{-a} & (\xi_k < \xi_s\ \&\ \omega_c < \omega_k < \omega_G),
\end{aligned} \right.~\label{eq:dispersion}
\end{equation}
which provides the dispersion relation with the exponent of $a > 1$ for the phonon-like vibrations of $\xi_k < \xi_s$.
By using the information of Eqs.~(\ref{eq:fractal}) and~(\ref{eq:dispersion}), we can predict the behavior of vDOS~\cite{Nakayama_1994} as
\begin{equation}
g(\omega)
\left\{ \begin{aligned}
&\propto \omega^{2}~(=\omega^{d-1}) & (\omega < \omega_\ast \sim \omega_c),\\
&\propto \omega^{D_f / a - 1}=\omega^{\tilde{d}-1} & (\omega_\ast \sim \omega_c < \omega < \omega_G),
\end{aligned} \right.~\label{eq:specdim}
\end{equation}
where $\tilde{d} = D_f/a$ is called the spectral dimension.
Note that for the case of $r > \xi_s$ and $\omega < \omega_\ast$, $D_f$ and $a$ correspond to $D_f = d=3$ and $a = 1$, which gives $\tilde{d} = d = 3$, i.e., the Debye scaling behavior.

We summarize the values of $D_f$, $a$, and $\tilde{d}=D_f/a$ in Table~\ref{table1}.
Figure~\ref{fig_vdos} plots by black lines the scaling of $g(\omega) \propto \omega^{\tilde{d}-1}$ in the plateau regime of $\omega_\ast < \omega < \omega_G$, which indeed confirms the validity of Eq.~(\ref{eq:specdim}) and the values of the spectral dimension $\tilde{d} = D_f /a$.
We can therefore understand the plateau of vDOS of gels in terms of the spectral dimension.

\begin{figure}[t]
\centering
\includegraphics[width=0.35\textwidth]{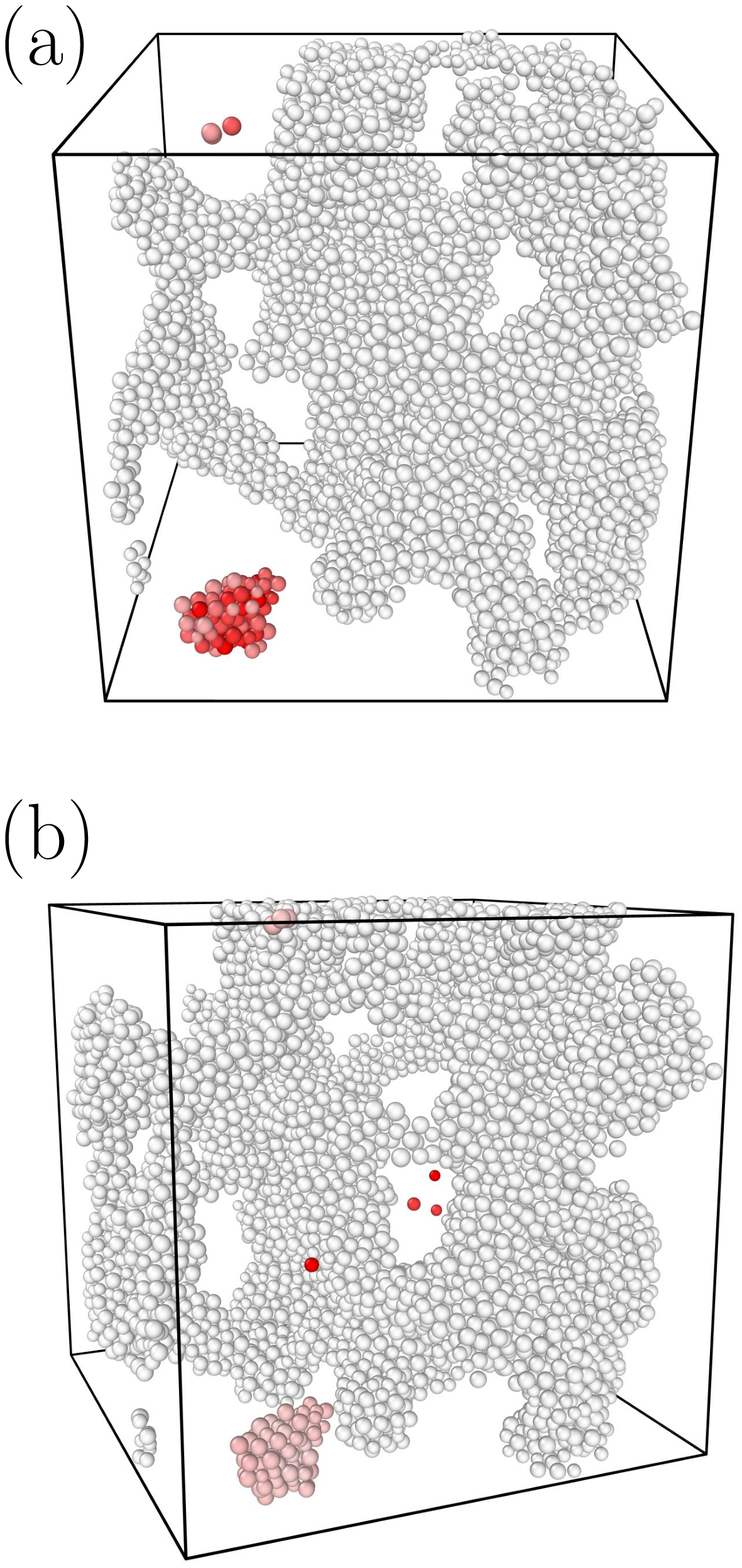}
\caption{\label{fig_vis_iso}
{Isolated vibrational states in the gel with $\rho =0.2$.}
(a) Isolated vibrations with finite frequency $\omega_k > \omega_G$ and (b) those with zero frequency $\omega_k = 0$~($\omega_k < 10^{-5}$).
In (a), for each particle $i$, we plot by red color summation of $|\mathbf{e}_i^k|$ over the modes with $\omega_k > \omega_G$ and $P_k < 10^{-2}$; $d_i = \sum_{k;\ \omega_k > \omega_G \ \&\ P_k < 10^{-2}} |\mathbf{e}_i^k|$.
In (b), we plot $d'_i = \sum_{k;\ \omega_k < 10^{-5}} |\mathbf{e}_i^k|$.
White color means zero value, while red color means finite $d_i$ or $d'_i$.
The number of particles is $N=10000$.
Only particles in an isolated cluster and isolated particles participate in these two vibrational states (a) and (b), with finite displacements.
}
\end{figure}

\subsection{Isolated vibrational states}~\label{subsec:isolated}
Finally, we note that localized vibrational modes exist with small values of $P^k < 10^{-2}$ and relatively high frequencies $\omega_k > \omega_G$ in the gel with $\rho = 0.2$, as recognized in panel (c) of Fig.~\ref{fig_character}.
To understand these modes, we calculate for each particle $i$ the summation of $|\mathbf{e}_i^k|$ over $\omega_k > \omega_G$ and $P_k < 10^{-2}$; $d_i = \sum_{k;\ \omega_k > \omega_G\ \&\ P_k < 10^{-2}} |\mathbf{e}_i^k|$.
Figure~\ref{fig_vis_iso}(a) plots by red color the spatial distribution of $d_i$ in the gel with $\rho =0.2$ and $N=10000$.
From the figure, we clearly recognize that all of the vibrating particles~(red particles) belong to a cluster that is isolated from the network structure, whereas the other particles do not at all participate in vibrations.
We therefore conclude that localized vibrations of $\omega_k > \omega_G$ are due to clusters isolated from the network structure of the gels.

In addition, as we also note in Sec.~\ref{subsect:method:vibration}, there are several modes with zero frequency $\omega_k = 0$~($\omega_k < 10^{-5}$) in the gels with $\rho = 0.2$ and $0.3$.
Similarly, for these zero modes, we calculate $d'_i = \sum_{k;\ \omega_k <  10^{-5}} |\mathbf{e}_i^k|$ for each particle $i$ and plot the spatial distribution of $d'_i$ in Fig.~\ref{fig_vis_iso}(b).
It is clear that zero modes originate from an isolated cluster of particles and isolated particles.
We therefore naturally conclude that isolated clusters and particles produce zero-frequency modes in the gel states.

\section{Discussion and conclusions}~\label{sect:conclusion}
In summary, we have studied the simplest model of particulate physical gels (and glass) composed of LJ particles at zero temperature and have provided a comprehensive understanding of their structural, mechanical, and vibrational properties.
\begin{enumerate}[(1)]
\item The gels show a sparse, heterogeneous, network-like structure, where clusters of glasses are connected to form the fractal structure of fractal dimension $2 < D_f < 3$.
As the density is lowered, the structure becomes sparser and more heterogeneous, which is captured by a growing characteristic length scale $\xi_s$.
\item Both the shear $G$ and bulk $K$ moduli of the gels significantly decrease with decreasing density.
The gels can become extremely soft, with elastic moduli orders of magnitude smaller than those of glasses.
In particular, the gels undergo the non-affine deformation even under isotropic bulk deformation and show rather small value of the bulk modulus.
\item Many low-frequency vibrational modes emerge, which form the characteristic plateau in the vDOS.
The vibrational states are changed at the onset frequency $\omega_\ast$ and the end frequency $\omega_G$ of the plateau~($\omega_\ast \ll \omega_G$).
At higher $\omega_G$, there is crossover between the phonon vibrations due to the elastic body at $\omega < \omega_G$ and the disordered vibrations due to the amorphous structure at $\omega > \omega_G$.
At $\omega < \omega_G$, the vibrations are phonon-like, showing crossover at the lower $\omega_\ast$ between those associated with a homogeneous elastic body at $\omega < \omega_\ast$ and those associated with a heterogeneous elastic body at $\omega > \omega_\ast$.
\item In the plateau regime of $\omega_\ast < \omega < \omega_G$, the fractal structure with $D_f$ plays an important role in the vibrational states.
The dispersion curve is described as $\omega_k \propto \xi_k^{-a}$ with $a>1$, which gives the vDOS $g(\omega) \propto \omega^{\tilde{d}-1}$ with the spectral dimension $\tilde{d} = D_f/a$.
\item Compared to the glasses, the BP is reduced and even absent in the gels, such that the vDOS smoothly converges to the Debye behavior at low frequencies below $\omega_\ast$.
Also, the QLV modes are suppressed in the gels.
These properties of BP and QLVs markedly contrast with those of glasses.
\item The abovementioned characteristic quantities show power-law dependences on the density $\rho$, such as $\xi_s \propto \rho^{-0.7}$, $G_0 \propto \rho^{2.8}$, and $\omega_\ast \propto \rho^{1.6}$.
These power-law scalings are closely related via $\omega_\ast \propto {c_T}/{\xi_s} \propto {\sqrt{G_0/\rho}}/\xi_s$, which establishes the relationship of structural, mechanical, and vibrational properties of the gels.
\end{enumerate}

In glasses, there is a characteristic frequency~(BP frequency) $\omega_\ast \sim \omega_G$ and associated length scale $\xi_\ast \propto c_T/\omega_\ast$~(or $\omega_\ast \propto c_T/\xi_\ast$)~\cite{Leonforte2005}.
$\omega_\ast$ and $\xi_\ast$ are the crossover points between the elastic body and the amorphous structural body.
Note that $\xi_\ast$ can be extracted from the elastic response to local deformation~\cite{Leonforte2005,Ellenbroek_2009,Lerner_2014} and global deformation~\cite{Karimi_2015,Mizuno_2019} or phonon transport properties~\cite{Mizuno_2018,Moriel_2019,Wang2_2019}.
In contrast to the glasses, the gels exhibit two characteristic frequencies, $\omega_\ast$ and $\omega_G$.
In the gels, the growing length $\xi_s$ of the static structure controls the $\omega_\ast$ as $\omega_\ast \propto c_T/\xi_s$ and separates two frequencies as $\omega_\ast \ll \omega_G$.
Note that the length $\xi_G \propto c_T/\omega_G$~(or $\omega_G \propto c_T/\xi_G$) (which corresponds to $\xi_\ast$ in the glasses) also exists in the gels, and since $\omega_G \gg \omega_\ast$, it is much smaller than $\xi_s$, $\xi_G \ll \xi_s$.
We therefore conclude that the present gels are multiscale, solid-state materials: (i) homogeneous elastic bodies at long lengths above $\xi_s$ and low frequencies below $\omega_\ast$, (ii) heterogeneous elastic bodies with fractal structures at intermediate lengths between $\xi_s$ and $\xi_G$ and frequencies between $\omega_\ast$ and $\omega_G$, and (iii) amorphous structural bodies at short lengths below $\xi_G$ and high frequencies above $\omega_G$.

In addition, it is remarkable that BP and QLVs are suppressed and even absent in the gels, which is a markedly contrasting situation with respect to glasses.
In glasses, repulsions are dominant between particle interactions, whereas attractive interactions play an important role in constructing network structures in gels~\cite{Tanaka_2004,Zaccarelli2007}.
It has been reported that in glasses, the strength of the BP and number of QLVs decrease with weakening repulsive interactions or strengthening attractive interactions between particles~\cite{Xu_2007,Lerner_2018,Karina_2021}.
From these observations, we speculate that in the gels, attractive forces may play a role in suppressing BP and QLVs.
This point should be clarified in detail in the future.

We note that the present gels can be considered porous glasses or aerogels such as silicate gels.
The most recent work~\cite{Niyogi_2021} in the context of porous glasses studied the mechanical properties of systems similar to the present ones.
They obtained density dependences of elastic moduli that are the same as those in Eq.~(\ref{eq.moduli}), and found scaling with the porosity that can be explained in a semi-empirical way~\cite{Phani_1987}.
Also, the recent simulations~\cite{Koeze_2018,Koeze_2020} demonstrated scalings with volume fraction in the jammed solids of sticky particles, where both the shear and bulk moduli vanish at the sticky jamming transition point.
In addition, a previous series of works~\cite{Courtens_1987,Courtens_1988,Vacher_1989,Vacher_1990,Anglaret_1994} performed scattering experiments, such as Brillouin, Raman, and inelastic neutron scattering, on silica aerogels and measured the vDOS~(and dispersion curve and line-width for acoustic excitations).
The characteristic plateau in the vDOS and crossover to the Debye behavior were observed, which are completely consistent with our simulation results.
This work also characterized vibrational modes in the plateau regime in terms of so-called fractons, which are highly localized vibrations on the fractal structure~\cite{percolation,Nakayama_1994,Alexander_1989}.
Further investigations are necessary to understand the relevance of our phonon vibrations to the heterogeneous structure with the fractons.

We also note that network-type glasses such as silicate glasses~\cite{Trachenko_2000} and glasses close to the jamming transition~\cite{Silbert_2005} also show a plateau in the vDOS.
The plateau in these systems is understood to originate from an isostatic nature~\cite{Wyart_2005,Wyart2_2005,Wyart_2006}, which is a different mechanism versus the case of gels, where the development of the plateau is simply associated with a reduction in global elastic moduli.
The scaled vDOS $g(\omega)/A_D\omega^2$ and BP diverge in glasses approaching the jamming transition~\cite{Mizuno_2017}, whereas $g(\omega)/A_D\omega^2$ smoothly converges to $1$ in the gels~(as explicitly plotted in Fig.~\ref{fig_scalevdos}).
In addition, vibrational modes in the plateaus are markedly different in nature between these glassy systems and the gels: they are anomalous modes with disordered vibrations in the glasses~\cite{Silbert_2009,Mizuno_2017}, whereas those in the gels show phonon-like vibrations.

Finally, we discuss the critical-like behaviors observed in the present particulate gels.
We found power-law scalings with density $\rho$ in the structure, elastic moduli, and vibrational states as $\xi_s \propto \rho^{-0.7}$, $G_0 \propto \rho^{2.8}$, $G_\text{ave} \propto K\propto \rho^{2.5}$~(or $G_0/G_A \propto \rho^{1.8}$, $G_\text{ave}/G_A \propto K/K_A\propto \rho^{1.5}$), and $\omega_\ast \propto \rho^{1.6}$.
This observation implies the existence of a critical phenomenon with the critical point at $\rho = 0$, where the length diverges, elastic moduli vanish, and the frequency goes to zero.
Note that our simulations show that the network structure breaks down at a density of $\rho = 0.1$.
In this work, we set the cut-off distance of the potential to be finite as $r_c =3.0$, which can result in the breakdown of the network.
We speculate that if $r_c$ is set to infinity, the gels retain the network structures as $\rho \to 0$, and the power-law scalings persist down to $\rho = 0$.
Previous experiments also reported scaling behaviors in the elastic moduli~\cite{Grant_1993,Krall_1998,Trappe2000,Prasad_2003} and the vibrational and acoustic properties~\cite{Courtens_1987,Courtens_1988,Vacher_1989,Vacher_1990,Anglaret_1994}.
These power-law scalings are consistent with predictions of rigidity percolation theory~\cite{Feng_1984,Kantor_1984,Feng2_1985}.
Note that the theory provides different values of critical points and exponents depending on the details of the models~\cite{Feng3_1985,Halperin_1985,Arbabi_1993,Sahimi_1993}.
Additionally, the jamming transition of glassy systems is well understood by using the mean-field approximation theory of rigidity percolation~\cite{Feng_1985,Wyart_2010,DeGiuli_2014}.
Further studies are necessary to understand the properties around the critical point of the gels.

\section*{Acknowledgments}
This work was supported by JSPS KAKENHI Grant Numbers 18H05225, 19K14670, 19H01812, 20H01868, and 20H00128.

\section*{Data availability}
The data that supports the findings of this study are available within the article and its supplementary material.

\bibliographystyle{apsrev4-1}
\bibliography{reference}

\clearpage
\section*{Supplementary Material}
\renewcommand{\thefigure}{S\arabic{figure}}
\renewcommand{\theequation}{S\arabic{equation}}
\setcounter{figure}{0}

\begin{figure}[t]
\centering
\includegraphics[width=0.475\textwidth]{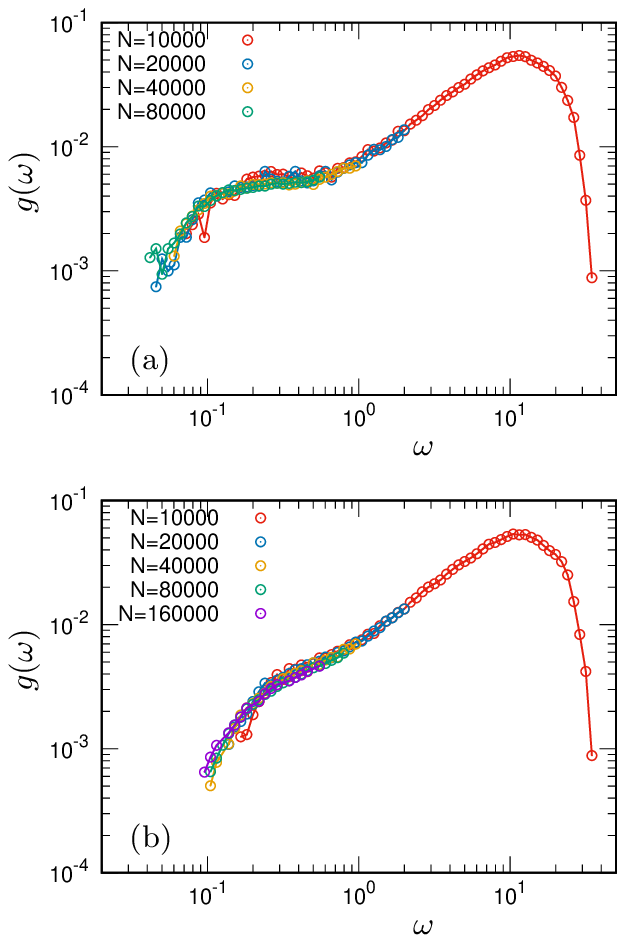}
\caption{\label{fig_dos_size}
{Vibrational density of states from different system sizes.}
$g(\omega)$ is plotted as a function of $\omega$ for different system sizes of $N=10000$ to $160000$.
The density is (a) $\rho=0.2$ and (b) $0.3$.
Data from different system sizes are smoothly connected as a function of $\omega$.
}
\end{figure}

\begin{figure}[t]
\centering
\includegraphics[width=0.475\textwidth]{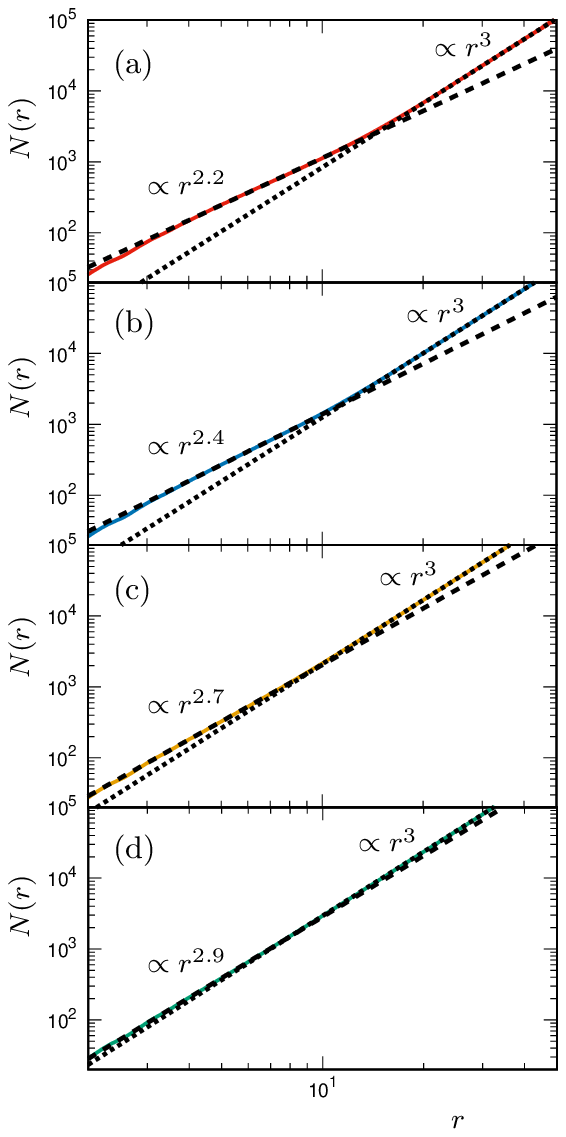}
\caption{\label{fig_intrdf2}
{Fractal dimension $D_f$ of $N(r) \propto r^{D_f}$~(at $r<\xi_s$).}
$N(r)$ is plotted as a function of $r$ for densities of (a) $\rho = 0.2$, (b) $0.3$, (c) $0.5$, and (d) $0.7$.
Data are the same as those presented in Fig.~\ref{fig_intrdf}.
Dotted lines indicate $\propto r^{3}$, whereas dashed lines indicate $\propto r^{D_f}$ with (a) $D_f=2.2$, (b) $2.4$, (c) $2.7$, and (d) $2.9$.
At approximately $r = \xi_s$, crossover occurs between $N(r) \propto r^3$ and $\propto r^{D_f}$.
}
\end{figure}

\begin{figure}[t]
\centering
\includegraphics[width=0.475\textwidth]{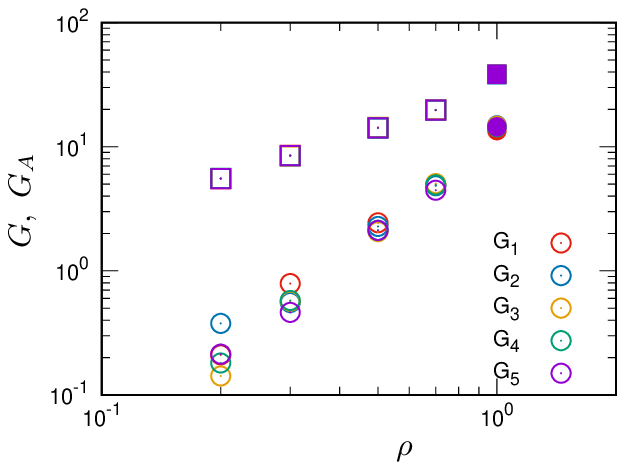}
\caption{\label{fig_moduli2}
{Five components of shear moduli.}
$G_1$, $G_2$, $G_3$, $G_4$, and $G_5$ are plotted as functions of $\rho$.
We also plot by squares the corresponding affine shear moduli.
Closed symbols represent data of the glass with $\rho=1.0$.
}
\end{figure}

\begin{figure*}[t]
\centering
\includegraphics[width=0.95\textwidth]{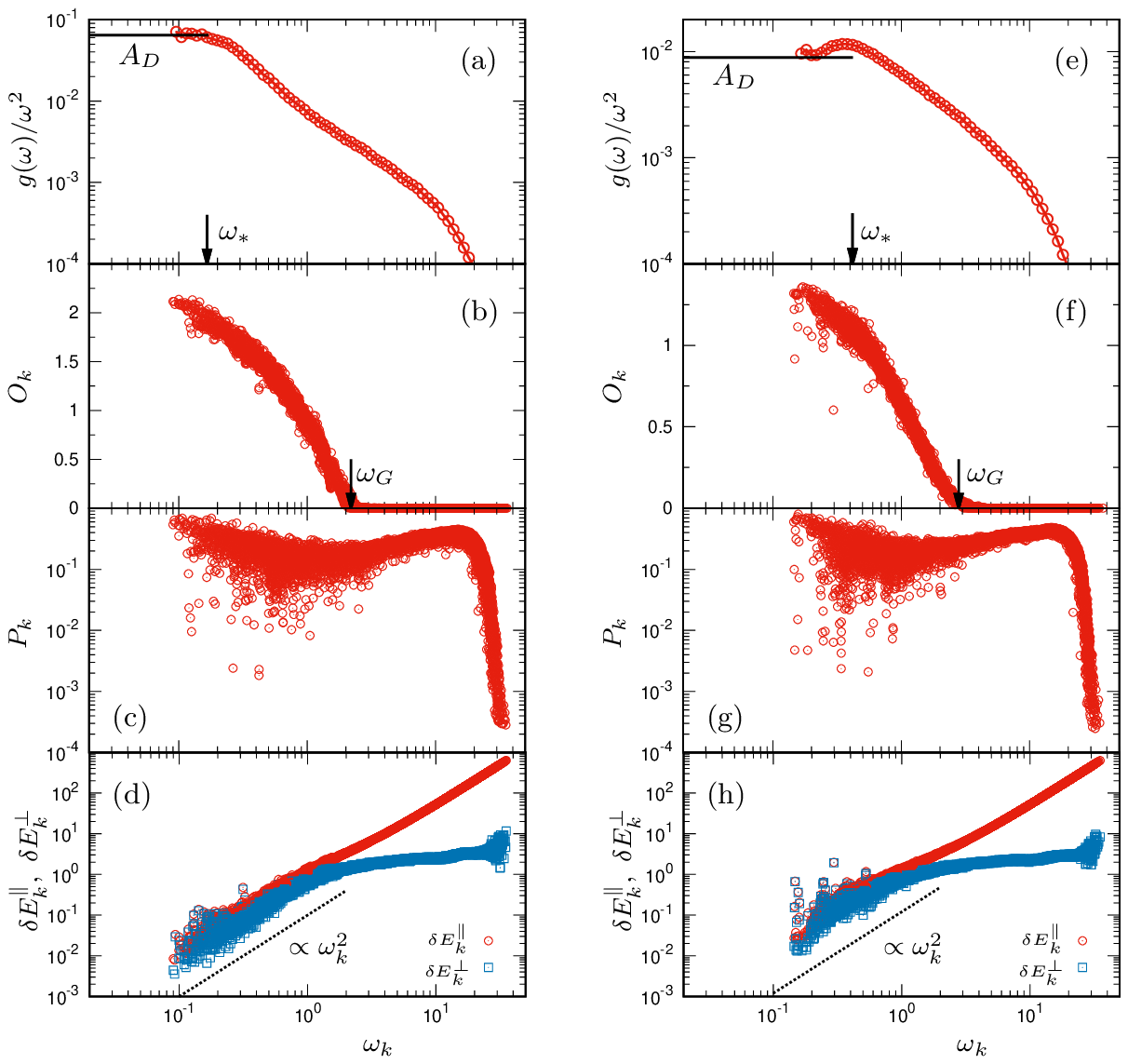}
\caption{\label{fig_character2}
{Characterization of vibrational modes.}
Plots of (a,e) $g(\omega)/\omega^2$, (b,f) $O_k$, (c,g) $P_k$, (d,h) $\delta E_k^{\parallel}$, and $\delta E_k^{\perp}$ of each mode $k$ as functions of $\omega_k$.
The left panels of (a) to (d) represent data of the gel with $\rho=0.3$, and the right panels of (e) to (h) are those of the gel with $\rho=0.5$.
The data of $g(\omega)/\omega^2$ are the same as those presented in Fig.~\ref{fig_vdos}(b).
The horizontal lines in panels~(a,e) present the Debye level $A_D$.
The arrows in (a,e) indicate the frequency $\omega_\ast$, whereas those in~(b,f) indicate $\omega_G$ at which $O_k$ becomes zero, $O_k \approx 0$.
}
\end{figure*}

\begin{figure}[t]
\centering
\includegraphics[width=0.475\textwidth]{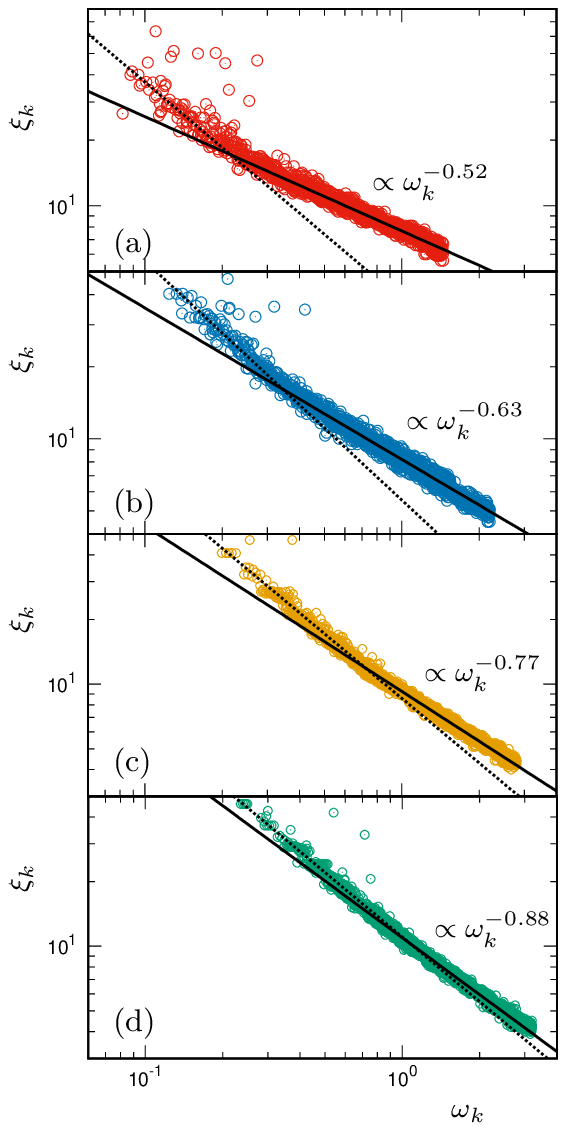}
\caption{\label{fig_length2}
{Exponent $1/a$ of $\xi_k \propto \omega_k^{-1/a}$~(at $\xi_k < \xi_s$).}
$\xi_k$ is plotted as a function of $\omega_k$ for densities of (a) $\rho = 0.2$, (b) $0.3$, (c) $0.5$, and (d) $0.7$.
Data are the same as those presented in Fig.~\ref{fig_length}.
Dotted lines indicate $\propto \omega_k^{-1}$, whereas solid lines indicate $\propto \omega_k^{-1/a}$ with (a) $1/a=0.52$, (b) $0.63$, (c) $0.77$, and (d) $0.88$.
At approximately $\omega_k = \omega_c$, crossover occurs between $\xi_k \propto \omega_k^{-1}$ and $\propto \omega_k^{-1/a}$.
}
\end{figure}

We report supplementary data, including the vDOS from different system sizes~(Fig.~\ref{fig_dos_size}), fractal dimension $D_f$ of $N(r) \propto r^{D_f}$~(Fig.~\ref{fig_intrdf2}), five components of shear moduli~(Fig.~\ref{fig_moduli2}), data of characterization of each vibrational mode for the gels with $\rho=0.3$ and $0.5$~(Fig.~\ref{fig_character2}), and exponent $1/a$ of $\xi_k \propto \omega_k^{-1/a}$~(Fig.~\ref{fig_length2}).

\end{document}